\documentclass[aps,prd,twocolumn,superscriptaddress,showpacs,nofootinbib,floatfix]{revtex4}

\usepackage{amsmath}
\usepackage{amssymb}
\usepackage{graphicx}

\renewcommand{\Vec}[1]{\boldsymbol{#1}}
\newcommand{\Deg}{^{\circ}}
\newcommand{\pT}{p_{T}}
\newcommand{\pTreco}{p_T^{\rm reco}}
\newcommand{\pTpy}{p_T^{\rm PY}}
\newcommand{\pTnlo}{p_T^{\rm NLO}}
\newcommand{\pTsum}{p_T^{\rm sum}}

\newcommand{\ALL}{A_{LL}}

\newcommand{\ALLcone}{A_{LL}^{\rm reco}}

\begin{document}
\title{
 Event Structure and Double Helicity Asymmetry in Jet Production from 
 Polarized $p+p$ Collisions at $\sqrt{s}$ = 200 GeV
}

\newcommand{\abilene}{Abilene Christian University, Abilene, Texas 79699, USA}
\newcommand{\banaras}{Department of Physics, Banaras Hindu University, Varanasi 221005, India}
\newcommand{\bnlcoll}{Collider-Accelerator Department, Brookhaven National Laboratory, Upton, New York 11973-5000, USA}
\newcommand{\bnlphys}{Brookhaven National Laboratory, Upton, New York 11973-5000, USA}
\newcommand{\caucr}{University of California - Riverside, Riverside, California 92521, USA}
\newcommand{\charlesczech}{Charles University, Ovocn\'{y} trh 5, Praha 1, 116 36, Prague, Czech Republic}
\newcommand{\ciae}{China Institute of Atomic Energy (CIAE), Beijing, People's Republic of China}
\newcommand{\cns}{Center for Nuclear Study, Graduate School of Science, University of Tokyo, 7-3-1 Hongo, Bunkyo, Tokyo 113-0033, Japan}
\newcommand{\colorado}{University of Colorado, Boulder, Colorado 80309, USA}
\newcommand{\columbia}{Columbia University, New York, New York 10027 and Nevis Laboratories, Irvington, New York 10533, USA}
\newcommand{\czechtech}{Czech Technical University, Zikova 4, 166 36 Prague 6, Czech Republic}
\newcommand{\dapnia}{Dapnia, CEA Saclay, F-91191, Gif-sur-Yvette, France}
\newcommand{\debrecen}{Debrecen University, H-4010 Debrecen, Egyetem t{\'e}r 1, Hungary}
\newcommand{\elte}{ELTE, E{\"o}tv{\"o}s Lor{\'a}nd University, H - 1117 Budapest, P{\'a}zm{\'a}ny P. s. 1/A, Hungary}
\newcommand{\fit}{Florida Institute of Technology, Melbourne, Florida 32901, USA}
\newcommand{\fsu}{Florida State University, Tallahassee, Florida 32306, USA}
\newcommand{\gsu}{Georgia State University, Atlanta, Georgia 30303, USA}
\newcommand{\hiroshima}{Hiroshima University, Kagamiyama, Higashi-Hiroshima 739-8526, Japan}
\newcommand{\ihepprot}{IHEP Protvino, State Research Center of Russian Federation, Institute for High Energy Physics, Protvino, 142281, Russia}
\newcommand{\illuiuc}{University of Illinois at Urbana-Champaign, Urbana, Illinois 61801, USA}
\newcommand{\instpasczech}{Institute of Physics, Academy of Sciences of the Czech Republic, Na Slovance 2, 182 21 Prague 8, Czech Republic}
\newcommand{\isu}{Iowa State University, Ames, Iowa 50011, USA}
\newcommand{\jinrdubna}{Joint Institute for Nuclear Research, 141980 Dubna, Moscow Region, Russia}
\newcommand{\kek}{KEK, High Energy Accelerator Research Organization, Tsukuba, Ibaraki 305-0801, Japan}
\newcommand{\kfki}{KFKI Research Institute for Particle and Nuclear Physics of the Hungarian Academy of Sciences (MTA KFKI RMKI), H-1525 Budapest 114, POBox 49, Budapest, Hungary}
\newcommand{\korea}{Korea University, Seoul, 136-701, Korea}
\newcommand{\kurchatov}{Russian Research Center ``Kurchatov Institute", Moscow, Russia}
\newcommand{\kyoto}{Kyoto University, Kyoto 606-8502, Japan}
\newcommand{\labllr}{Laboratoire Leprince-Ringuet, Ecole Polytechnique, CNRS-IN2P3, Route de Saclay, F-91128, Palaiseau, France}
\newcommand{\lawllnl}{Lawrence Livermore National Laboratory, Livermore, California 94550, USA}
\newcommand{\losalamos}{Los Alamos National Laboratory, Los Alamos, New Mexico 87545, USA}
\newcommand{\lpc}{LPC, Universit{\'e} Blaise Pascal, CNRS-IN2P3, Clermont-Fd, 63177 Aubiere Cedex, France}
\newcommand{\lund}{Department of Physics, Lund University, Box 118, SE-221 00 Lund, Sweden}
\newcommand{\muenster}{Institut f\"ur Kernphysik, University of Muenster, D-48149 Muenster, Germany}
\newcommand{\myongji}{Myongji University, Yongin, Kyonggido 449-728, Korea}
\newcommand{\nagasaki}{Nagasaki Institute of Applied Science, Nagasaki-shi, Nagasaki 851-0193, Japan}
\newcommand{\newmex}{University of New Mexico, Albuquerque, New Mexico 87131, USA }
\newcommand{\nmsu}{New Mexico State University, Las Cruces, New Mexico 88003, USA}
\newcommand{\ornl}{Oak Ridge National Laboratory, Oak Ridge, Tennessee 37831, USA}
\newcommand{\orsay}{IPN-Orsay, Universite Paris Sud, CNRS-IN2P3, BP1, F-91406, Orsay, France}
\newcommand{\peking}{Peking University, Beijing, People's Republic of China}
\newcommand{\pnpi}{PNPI, Petersburg Nuclear Physics Institute, Gatchina, Leningrad region, 188300, Russia}
\newcommand{\riken}{RIKEN Nishina Center for Accelerator-Based Science, Wako, Saitama 351-0198, Japan}
\newcommand{\rikjrbrc}{RIKEN BNL Research Center, Brookhaven National Laboratory, Upton, New York 11973-5000, USA}
\newcommand{\rikkyo}{Physics Department, Rikkyo University, 3-34-1 Nishi-Ikebukuro, Toshima, Tokyo 171-8501, Japan}
\newcommand{\saispbstu}{Saint Petersburg State Polytechnic University, St. Petersburg, Russia}
\newcommand{\saopaulo}{Universidade de S{\~a}o Paulo, Instituto de F\'{\i}sica, Caixa Postal 66318, S{\~a}o Paulo CEP05315-970, Brazil}
\newcommand{\seoulnat}{Seoul National University, Seoul, Korea}
\newcommand{\stonybrkc}{Chemistry Department, Stony Brook University, SUNY, Stony Brook, New York 11794-3400, USA}
\newcommand{\stonycrkp}{Department of Physics and Astronomy, Stony Brook University, SUNY, Stony Brook, New York 11794-3400, USA}
\newcommand{\subatech}{SUBATECH (Ecole des Mines de Nantes, CNRS-IN2P3, Universit{\'e} de Nantes) BP 20722 - 44307, Nantes, France}
\newcommand{\tenn}{University of Tennessee, Knoxville, Tennessee 37996, USA}
\newcommand{\titech}{Department of Physics, Tokyo Institute of Technology, Oh-okayama, Meguro, Tokyo 152-8551, Japan}
\newcommand{\tsukuba}{Institute of Physics, University of Tsukuba, Tsukuba, Ibaraki 305, Japan}
\newcommand{\vandy}{Vanderbilt University, Nashville, Tennessee 37235, USA}
\newcommand{\waseda}{Waseda University, Advanced Research Institute for Science and Engineering, 17 Kikui-cho, Shinjuku-ku, Tokyo 162-0044, Japan}
\newcommand{\weizmann}{Weizmann Institute, Rehovot 76100, Israel}
\newcommand{\yonsei}{Yonsei University, IPAP, Seoul 120-749, Korea}
\affiliation{\abilene}
\affiliation{\banaras}
\affiliation{\bnlcoll}
\affiliation{\bnlphys}
\affiliation{\caucr}
\affiliation{\charlesczech}
\affiliation{\ciae}
\affiliation{\cns}
\affiliation{\colorado}
\affiliation{\columbia}
\affiliation{\czechtech}
\affiliation{\dapnia}
\affiliation{\debrecen}
\affiliation{\elte}
\affiliation{\fit}
\affiliation{\fsu}
\affiliation{\gsu}
\affiliation{\hiroshima}
\affiliation{\ihepprot}
\affiliation{\illuiuc}
\affiliation{\instpasczech}
\affiliation{\isu}
\affiliation{\jinrdubna}
\affiliation{\kek}
\affiliation{\kfki}
\affiliation{\korea}
\affiliation{\kurchatov}
\affiliation{\kyoto}
\affiliation{\labllr}
\affiliation{\lawllnl}
\affiliation{\losalamos}
\affiliation{\lpc}
\affiliation{\lund}
\affiliation{\muenster}
\affiliation{\myongji}
\affiliation{\nagasaki}
\affiliation{\newmex}
\affiliation{\nmsu}
\affiliation{\ornl}
\affiliation{\orsay}
\affiliation{\peking}
\affiliation{\pnpi}
\affiliation{\riken}
\affiliation{\rikjrbrc}
\affiliation{\rikkyo}
\affiliation{\saispbstu}
\affiliation{\saopaulo}
\affiliation{\seoulnat}
\affiliation{\stonybrkc}
\affiliation{\stonycrkp}
\affiliation{\subatech}
\affiliation{\tenn}
\affiliation{\titech}
\affiliation{\tsukuba}
\affiliation{\vandy}
\affiliation{\waseda}
\affiliation{\weizmann}
\affiliation{\yonsei}
\author{A.~Adare} \affiliation{\colorado}
\author{S.~Afanasiev} \affiliation{\jinrdubna}
\author{C.~Aidala} \affiliation{\columbia}
\author{N.N.~Ajitanand} \affiliation{\stonybrkc}
\author{Y.~Akiba} \affiliation{\riken} \affiliation{\rikjrbrc}
\author{H.~Al-Bataineh} \affiliation{\nmsu}
\author{J.~Alexander} \affiliation{\stonybrkc}
\author{K.~Aoki} \affiliation{\kyoto} \affiliation{\riken}
\author{L.~Aphecetche} \affiliation{\subatech}
\author{R.~Armendariz} \affiliation{\nmsu}
% ...
\collaboration{PHENIX Collaboration} \noaffiliation

\date{\today}

\begin{abstract}

We report on the event structure and double helicity asymmetry 
($A_LL$) of jet production in longitudinally polarized p+p 
collisions at $\sqrt{s}$=200 GeV.  Photons and charged particles 
were measured by the PHENIX experiment at midrapidity 
$|\eta| < 0.35$ with the requirement of a high-momentum 
($>2$ GeV/$c$) photon in the event. Event structure, such as 
multiplicity, $p_T$ density and thrust in the PHENIX acceptance, 
were measured and compared with the {\sc pythia} event generator 
and the {\sc geant} detector simulation.  The shape of jets and the 
underlying event were well reproduced at this collision energy. 
For the measurement of jet $A_{LL}$, photons and charged 
particles were clustered with a seed-cone algorithm to obtain 
the cluster $p_T$ sum ($p_T^{\rm reco}$). The effect of detector 
response and the underlying events on $p_T^{\rm reco}$ was 
evaluated with the simulation. The production rate of 
reconstructed jets is satisfactorily reproduced with the NLO 
pQCD jet production cross section. 
For $4 < p_T^{\rm reco} < 12$ GeV/$c$ with an average beam 
polarization of $\langle P \rangle = 49\%$ we measured $A_{LL} = 
-0.0014 \pm 0.0037^{\rm stat}$ at the lowest $p_T^{\rm reco}$ 
bin (4--5 GeV/$c$) and $-0.0181 \pm 0.0282^{\rm stat}$ at the 
highest $p_T^{\rm reco}$ bin (10--12 GeV/$c$) with a beam 
polarization scale error of 9.4\% and a $\pT$ scale error of 
10\%. Jets in the measured $p_T^{\rm reco}$ range arise 
primarily from hard-scattered gluons with momentum fraction 
$0.02 < x < 0.3$ according to {\sc pythia}. The measured 
$A_{LL}$ is compared with predictions that assume various 
$\Delta G(x)$ distributions based on the GRSV parameterization. 
The present result imposes the limit $-1.1 < \int_{0.02}^{0.3}dx 
\Delta G(x, \mu^2 = 1 {\rm GeV}^2) < 0.4$ at 95\% confidence 
level or $\int_{0.02}^{0.3}dx \Delta G(x, \mu^2 = 1 {\rm 
GeV}^2) < 0.5$ at 99\% confidence level.

\end{abstract}


\maketitle

\section{Introduction}

The motivation of this measurement is to understand 
the spin structure of the proton, particularly
the contribution of the gluon spin ($\Delta G$) to the proton spin.
The proton spin can be represented as
\begin{equation}
\frac{1}{2}_{proton} = \frac{1}{2}\sum_f \Delta q_f + \Delta G + L_q + L_g,
\end{equation}
where $\Delta G$ is the gluon spin, i.e.~the integral of the polarized 
gluon distribution function, $\Delta G = \int_0^1 dx \Delta G(x)$, 
$\sum\Delta q$ is the quark spin, and $L_q$ and $L_g$ are the orbital 
angular momenta of quarks and gluons in the proton. It was found by the 
EMC experiment at CERN in 1987 that the quark spin contribution to the 
proton spin is only 
$(12 \pm 9 \pm 14)\%$~\cite{Ashman:1987hv,Ashman:1989ig}. 
After the EMC experiment many deep inelastic scattering (DIS) experiments 
have been carried out to measure $\sum\Delta q$ more precisely. The 
recent analysis by the HERMES experiment~\cite{Airapetian:2007mh} reported 
that $\sum\Delta q = 0.330 \pm 0.011 ({\rm theo.}) \pm 0.025 ({\rm exp.}) 
\pm 0.028 ({\rm evol.})$ at a hard-scattering scale $\mu^2 \sim 5$ 
GeV$^2$, which is only about 30\% of the proton spin.  
Consequently, the majority of the proton spin should be carried by the 
remaining components.

Jet production from
longitudinally-polarized $p+p$ collisions is suited for the
measurement of $\Delta G$ because gluon-involved scatterings, such
as $q+g \to q+g$ or $g+g \to g+g$, dominate the cross section.
The double helicity asymmetry 
\begin{equation}
\ALL \equiv \frac{\sigma_{++} - \sigma_{+-}}{\sigma_{++} + \sigma_{+-}},
\end{equation}
is the asymmetry in cross section between two beam helicity states.
In the $\ALL$ measurement, 
many systematic errors cancel out so that high precision can be achieved.

Another motivation of this measurement is to study 
the event structure of $p+p$ collisions.
A high-energy $p+p$ collision produces not only hard scattered
partons but also many particles that originate from soft interactions which we call the `underlying event'.
The {\sc pythia} event generator phenomenologically models the underlying event
on the Multi-Parton Interaction (MPI) scheme~\cite{pythiaweb},
and can reproduce the event structure of $p+\bar{p}$ collisions measured by
the CDF experiment at $\sqrt{s}$ = 1.8 TeV~\cite{Field:2005qt}.
We present measurements of event structure at lower collision energy, 
$\sqrt{s}$ = 200 GeV, and compare them with
those simulated by {\sc pythia} in order to examine the validity of
the {\sc pythia} MPI scheme.
One of the goals of the PHENIX experiment at the Relativistic Heavy
Ion Collider (RHIC) is the determination of
$\Delta G$.
PHENIX has published results on single particle production;
the $\ALL$ of $\pi^0$ production was reported 
in~\cite{Adare:2008qb,Adare:2008px}.
This paper reports a measurement of jet production.
For $\Delta G$, it is valuable to determine the parton kinematics 
following the collision in order to better control the $x$ range. 
In this work we reconstruct jets, observing a larger fraction of the 
parton\'s 
momentum. This allows improved reconstruction of the original parton 
kinematics and better statistical accuracy for higher $x$ gluons.
Since $\pi^0$'s in $p+p$ collisions are produced via
jet fragmentation,
the measurements of jet and $\pi^0$ with same data set
have a statistical overlap.  The size of the overlap was estimated to be 
40-60\% depending on the jet $\pT$.  Even in such overlapped events,
measured $\pT$ of jets does not correlate with that of $\pi^0$s,
and thus the two measurements have an independent sensitivity on $x$.
The fraction of $q+g$ subprocess is larger than $q+q$ and $g+g$ subprocesses
in the present jet measurement, making it sensitive to
the sign of $\Delta G$.
The STAR experiment at RHIC is also measuring inclusive jets to determine 
$\Delta G$~\cite{Abelev:2007vt}.
These measurements have different types of systematic uncertainties
and thus one can provide a systematic check for the other.

The remainder of this paper is organized as follows.
In Section II,
the parts of the PHENIX detector that is relevant to the jet
measurement are described.
In Section III, 
analysis methods such as particle clustering and simulation studies
are discussed.
In Section IV, 
results on event structure, jet production rate and 
beam-helicity asymmetries are shown.

\section{Experimental Setup}

\begin{figure}[bthp] 
\includegraphics[width=1.0\linewidth]{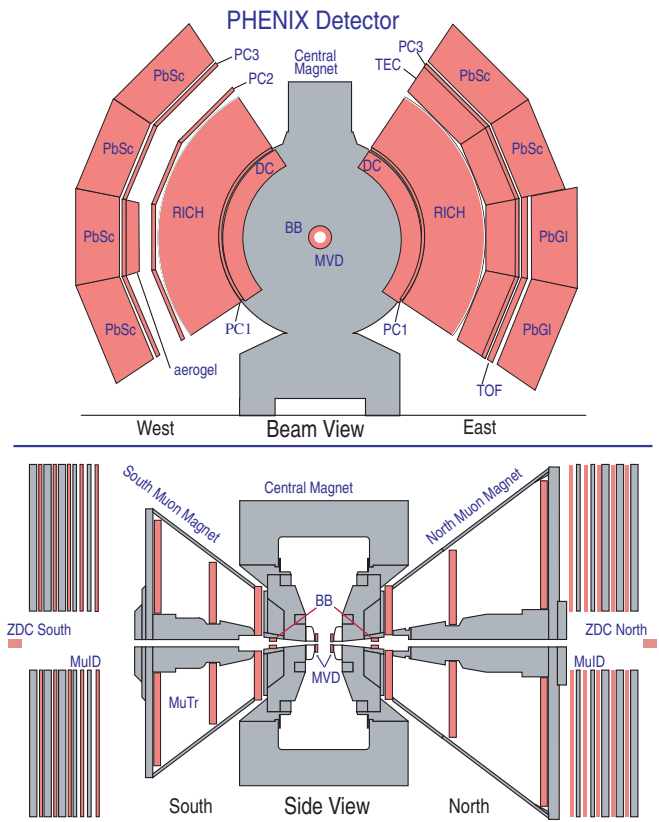}
\caption{ (color online) PHENIX detector. }
\label{fig:phenix2004}
\end{figure}

The PHENIX detector~\cite{Adcox:2003zm} can be grouped into three parts;
the Inner Detectors, the Central Arms and the Muon Arms.
The schematic drawing of the PHENIX detector is shown in
Fig.~\ref{fig:phenix2004}.
In this measurement,
the Central Arms were used to detect photons and charged particles in
jets, and the Inner Detectors to obtain the collision vertex and beam
luminosity.

\subsection{Inner detectors}

The Inner Detectors include the Beam-Beam Counters (BBC) and the
Zero-Degree Calorimeters (ZDC).

The BBC is composed of two identical sets of counters placed at
both the north and south sides of the collision
point with a 144 cm distance~\cite{Allen:2003zt}.
Each counter is composed of 64 sets of PMT plus a 3-cm quartz \v{C}erenkov
radiator.
The BBC covers a pseudorapidity of $3.0 < \left| \eta \right| < 3.9$ over
the full azimuth.
The BBC measures the number of charged particles in forward and backward
regions to determine the collision time, collision $z$-vertex, and beam
luminosity.
The timing and $z$-vertex resolution in $p+p$
collisions are about 100 ps and 2 cm, respectively.

The ZDC is comprised of two sets of hadronic calorimeters placed at the north 
and south sides of the collision point with a 18 m
distance~\cite{Adler:2000bd}.
It covers a 10 cm $\times$ 10 cm area perpendicular to the beam
direction, which corresponds to 2.8 mrad when viewed from the collision
point.
It consists of alternating layers of tungsten absorbers and sampling
fibers, and is 150 radiation lengths and 5.1 interaction lengths in depth.
It measures neutrons in forward and backward regions and is used as
a local polarimeter which assures that the beam polarization is
correctly longitudinal or transverse at the interaction region by
observing the left-right asymmetry in the $\vec{p}+p \to {\rm neutron}+X$
scattering cross section~\cite{Bazilevsky:2006vd,Adare:2007dg}.

\subsection{Central Arms}

The Central Arms consist of a tracking system and an electromagnetic
calorimeter (EMCal).
Pad chambers (PC) and drift chambers (DC) were used to detect charged
particles in jets, and the EMCal was used to detect photons in jets.

The EMCal system~\cite{Aphecetche:2003zr} 
is located at a distance of 5 m from the interaction point.
The system consists of four sectors in each of the East and West Arms,
and each sector has a size of 2$\times$4 m$^2$.
The system is composed of two types of calorimeter, lead scintillator
(PbSc) and lead glass (PbGl).
One PbSc module has a size of 5.5$\times$5.5$\times$37.5 cm$^3$
corresponding to 18.0 radiation lengths.
One PbGl module has a size of 4.0$\times$4.0$\times$40.0 cm$^3$
corresponding to 14.4 radiation lengths.
The energy resolution is $\sim 7\%$ at $E=1$ GeV.

The DC system~\cite{Adcox:2003zp} is located in the region from 2 to 2.4
m from the interaction point to measure the position and momentum of charged
particles.
The DC system consists of one frame in each of the East and West Arms.
Each chamber has a size of 2.5 m$\times$90$\Deg$ in $z$-$\phi$ direction
with cylindrical shape, and is composed of 80 sense planes 
with a 2-2.5 cm drift space in the $\phi$ direction.
Each sense plane has 24 wires, 
which precisely measure $r$-$\phi$ position, 
and 16 tilted wires, which measure $z$ position.

The PC system~\cite{Adcox:2003zp} is composed of multi-wire proportional
chambers in three separate layers, which are called PC1, PC2 and
PC3, of the Central Arms tracking system.
The PC1 is located behind the DC and is used for pattern recognition
together with the DC by providing the $z$ coordinate.
The PC1 consists of a single plane of anode and field wires lying in a
gas volume between two cathode planes.
One cathode is segmented into pixels 
with a size of $\sim 8.5 \times 8.5$ mm$^2$, and
signals from the pixels are read out.

Charged particle tracks are reconstructed using the information from the
DC and the PC1~\cite{Mitchell:2002wu}.
The magnetic field between the collision vertex and the DC is axial, and
thus bends particles in the $x$-$y$ plane.
The field is so weak at the outer area from the DC that particle
tracks can be assumed to be straight.
A track reconstruction is performed in the DC first, and
then reconstructed tracks are associated with hits in the PC1.
The momentum resolution is given by
$\sigma_p / p~(\%) = 1.3 \cdot p~({\rm GeV}/c) \oplus 1.0$ for pions.

\subsection{Trigger} \label{sec:trigger}

The PHENIX experiment has various trigger configurations to efficiently
select many type of interesting rare events.
This measurement required the coincidence of two triggers;
a minimum bias (MB) trigger issued by the BBC, and
a high-energy photon trigger issued by the EMCal.

The MB trigger in $p+p$ collisions requires one charged particle
in both the north and south sides of the BBC.
The reconstructed $z$-vertex is required to be within $\pm\sim30$ cm.
The efficiency, $f_{\rm MB}$, of the MB trigger for high-$\pT$ QCD scatterings 
such as jet production is $0.784 \pm 0.020$, which has been determined with 
the ratio of $\pi^0$ yields with and without the MB trigger requirement.

The high-energy photon trigger is fired when the sum of energy deposits
in $4\times4$ EMCal modules 
($\Delta \phi \simeq \Delta \eta \simeq 0.04$)
is above a threshold, $\sim 1.4$ GeV, which varies by $\sim 0.2$ GeV 
area-by-area due to the variations of gain and threshold between EMCal modules.
Each $4\times4$ area overlaps with others, and thus even when a photon hits
the edge of a $4\times4$ area the next overlapped $4\times4$ area can
gather all energy of the photon.
The trigger efficiency is almost flat and close to unity 
above $E \sim 2$ GeV except masked areas due to noise in the trigger electronics.

\section{Analysis Methods}

\subsection{Outline}

This analysis used 2.3 pb$^{-1}$ of data that were taken with 
the MB + high-energy-photon trigger in 2005.
In addition, $\sim$0.3 pb$^{-1}$ of data that were taken with the MB trigger
alone were used for systematic error studies.
Photons were detected with the EMCal, and
charged particles were detected with the DC and PC1.
Measured particles in each PHENIX Central Arm were clustered using a cone method 
to form a `reconstructed jet' and its transverse momentum ($\pTreco$).
Because of the finite size of the acceptance ($|\eta| < 0.35$),
the cone size for the particle clustering were set to 0.3 at maximum.
This is smaller than the typical cone size, 0.7 raising two issues:
First, a jet in an NLO calculation is usually defined with the same cone size
and compared with the measured jet,
but this is optimum when both jet energy and cone size are large 
since the jet spread due to hadronization becomes significant with
small jet energy and cone size.
Second, such a small cone is more sensitive to quark jets
than gluon jets since gluon jets are broader and softer than quark jets.
Because of the situation described above,
the theory calculation and the simulation evaluations
have been organized as follows.

The cross section and the $\ALL$ of inclusive jet production 
were calculated as a function of jet transverse momentum ($\pTnlo$)
within the framework of a next-to-leading-order perturbative QCD (NLO pQCD).
This calculation predicted various $\ALL$'s 
by assuming various $\Delta G(x)$ distributions.

A simulation with the {\sc pythia} event generator~\cite{pythiaweb} and 
the {\sc geant} detector simulation package~\cite{geantweb} was performed 
to understand the effects of the detector response, the underlying events
and the jet-definition difference between the measurement and 
the theory calculation.
{\sc pythia} simulates parton-parton hard scatterings in $p+p$ collisions
at leading order (LO) in $\alpha_s$
with phenomenological initial and final-state radiation and hadronization.
{\sc geant} simulates the acceptance and response of the PHENIX detector.
We define a jet at the partonic level in {\sc pythia}.
The effect of the detector response and the underlying events
was evaluated as the statistical relation between the jets defined in {\sc pythia}
and the reconstructed jets.
We assume $\pTpy = \pTnlo$
within an uncertainty that will be explained in a later section, 
and then we obtained the relation between
the NLO calculation and the measurement.

To confirm that the simulation reproduces well the real data in terms of
event structure, namely spatial distribution of particles in an event,
quantities sensitive to event structure were measured.
Those include particle multiplicity, transverse-momentum density, 
thrust distribution and jet-production rate.
A comparison was made between the real data and the simulation output.

We derive the predictions of the measured $\ALL$ 
by converting the NLO calculation with the relation between $\pTnlo$ and
$\pTreco$.
A $\chi^2$ test between the measured and predicted $\ALL$'s was performed
to determine the most-probable $\Delta G$.

\begin{figure}[bthp]
\includegraphics[width=1.0\linewidth]{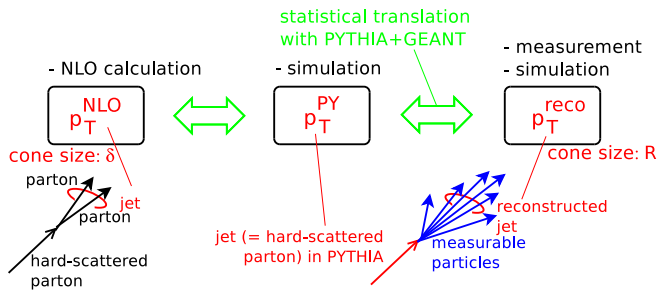}
\caption{ (color online) Relations between the jets defined in this measurement. }
\label{fig:pT_relation}
\end{figure}

The definitions and relations of jets in this measurement are
summarized in Tab.~\ref{tab:jet_def} and Fig.~\ref{fig:pT_relation}.

\begin{table}[bthp] 

\caption{ Definitions of jets adopted in this measurement.}
\label{tab:jet_def}
\begin{ruledtabular} \begin{tabular}{p{.34\linewidth}p{.58\linewidth}}
Reconstructed jet\newline
\hspace*{2ex}($\pTreco$)
  & Hadronic jet made with measurable particles after hadronization\newline
    with a cone size of $R = 0.3$. \\
jet in {\sc pythia}\newline
\hspace*{2ex}($\pTpy$)
  & Partonic jet in {\sc pythia}\newline
    without cone. \\
\hline
jet in NLO calculation\newline
\hspace*{2ex}( $\pTnlo$)
  & Partonic jet in NLO pQCD calc.\newline
    with a cone size of $\delta = 1.0$.  \\
\end{tabular} \end{ruledtabular}
\end{table}

\subsection{Particle Clustering with Cone Method} \label{sec:clustering}

A jet in one PHENIX Central Arm is constructed with photons and charged
particles detected with the EMCal, the DC and the PC1 of the Central
Arm.
A seed-cone algorithm, described below, is used for the cluster finding.

\subsubsection{Event and particle selections}

To select the energy region where the efficiency of the high-$\pT$
photon trigger is in the plateau,
at least one photon with $\pT > 2.0$ GeV$/c$
is required in each event.
This requirement causes a bias towards jets that include mostly high-$\pT$
$\pi^0$, $\eta$, etc.~or radiated photons.

To collect photons from all EMCal hits, 
a $\pT$ cut, a charged track veto, and an EMCal shower shape cut
were applied.
The $\pT$ cut required the $\pT$ of each EMCal hit to be $>0.4$ GeV$/c$
in order to eliminate hits likely to be dominated by electronics noise in the detector.
It also eliminates charged hadron hits
because the measured energy of minimum ionization particles by PbSc 
peaks at 0.25 GeV.
and that of $\pi^\pm$ with momentum of 1 GeV/$c$ 
in the PbGl result in a distribution
peaked around 0.4 GeV, with a broad tail to lower energy.
The charged track veto reduces charged particle contamination 
by checking whether each EMCal hit has a matched charged
track within $3\sigma$ of their position resolutions.
The shower shape cut reduces hadron contamination
by comparing the fraction of energy deposits in every EMCal module
of a hit with the fraction predicted by a model of shower shape.
This cut eliminates half of hadron hits and statistically 1\% of
photon hits.
These cuts made the contamination of charged and neutral hadrons
negligible.

All charged particles detected with the DC and the PC1
were required to have $\pT$ ranging from 0.4 to 4.0 GeV/$c$.
Below the lower limit, 
the acceptance is strongly distorted due to a large bending angle and thus
becomes shifted from that of photons.
The upper limit eliminates fake high-$\pT$ tracks which originate from
low-$\pT$ particles that are produced from a decay or a conversion in the magnetic field.
Note that this limit causes 
a bias towards jets that include fewer charged particles.

\subsubsection{Cluster finding algorithm}

All particles that satisfy the experimental cuts in one arm were used
as a seed in cluster finding.
Starting with the momentum direction of a seed particle as a temporary
cone axis, we calculated the next temporary cone axis with particles which
are in the cone.
The distance between the cone axis $(\eta^C, \phi^C)$ and the momentum
direction of each particle $(\eta^i, \phi^i)$ is defined as
\begin{equation}
R^i \equiv \sqrt{(\eta^i - \eta^C)^2 + (\phi^i - \phi^C)^2}.
\end{equation}
The cone radius $R$ was set to 0.3, which was about a half of the
$\eta$ acceptance of the detector.
The next temporary cone axis $\vec{e}_{\rm next}$ is calculated as a
vector sum of momenta of particles in the cone:
\begin{equation}
\vec{e}_{\rm next} \equiv
  \frac{\vec{p}^{\rm \ reco}}{\left| \vec{p}^{\rm \ reco} \right|}
\ \ , \ \ \ 
\vec{p}^{\rm \ reco} \equiv
  \sum_{i\in{\rm cone}} \vec{p}_i.
\end{equation}
This procedure was iterated until the temporary cone axis became stable.

The cluster finding is done with all seed particles, and then each seed
particle has one cone and some cones can be the same or overlapped.
The cone which has the largest $\pTreco$ in an event is used in
the event.
For measurements of event structure we also define the sum of momenta of 
all particles in one arm:
\begin{equation}
\vec{p}^{\rm \ sum} \equiv \sum_{i\in{\rm arm}} \vec{p}_i.
\end{equation}

An evaluation of $\pTreco$ without seed has been done using a part of
the statistics in order to check the effect of the use of a seed.
Every direction in the ($\eta$, $\phi$) space with 
a step of $\delta\eta = \delta\phi = 0.01$ within the Central Arm acceptance
has been used as an initial cone direction in each event.
All steps except the choice of the initial cone directions is the same as the original algorithm.
The yield of reconstructed jets with the seedless method was larger than that
with the seed method by $\sim$20\% at $\pTreco = 4$ GeV/$c$, 
$\sim$10\% at $\pTreco = 8$ GeV/$c$ and 
$\sim$5\% at $\pTreco = 12$ GeV/$c$.
This deviation is compensated in the relation between $\pTreco$ and $\pTpy$
estimated with the simulation, 
and therefore the $\pTreco$ difference between the two methods of 
cluster finding is smaller than the deviation above.

\subsection{Simulation Study} \label{sec:simulation}

\subsubsection{Simulation settings}

The {\sc pythia} version 6.220 was used.
Only QCD high-$\pT$ processes were generated 
by setting the process switch (``MSEL'') to 1 and 
the lower cutoff of partonic transverse momentum (``CKIN(3)'') to 1.5 GeV/$c$.
The parameter modification reduces the time for event generation
and does not affect any physics results in the measured $\pT$ region,
as it has been confirmed by comparing $\pTreco$ distribution etc.~to 
those without the parameter modification.
We call a {\sc pythia} simulation with these conditions `{\sc pythia} default'.
Hadron-hadron collisions have a so-called `underlying event', 
which comes from the breakup of the incident nucleons.
The {\sc pythia} simulation reproduces the underlying event with the Multi-Parton Interaction (MPI) mechanism.
The CDF experiment at the Tevatron showed that
the {\sc pythia} simulation did not reproduce the event structure well and 
modeled a set of tuned parameters called 
`tune A'~\cite{Field:2005qt,Field:web_page}.
Modified or important parameters are listed in Tab.~\ref{tab:pythia_mpi_param}.

\begin{table}[bthp] 
\caption{Important or modified (Used) parameters in the {\sc pythia} MPI 
setting.}
\label{tab:pythia_mpi_param}
\begin{ruledtabular} \begin{tabular}{cccp{.5\linewidth}}
  Parameter & Default & Used & Note \\
\hline
  MSTP(81)  & 1       & 1    & MPI master switch. \\
  MSTP(82)  & 1       & 4    & double-Gaussian matter distribution used. \\
  PARP(82)  & 1.9     & 2.0  & turn-off $\pT$ for MPI at the reference energy scale PARP(89) \\
  PARP(83)  & 0.5     & 0.5  & the fraction of the core Gaussian matter to total hadronic matter \\
  PARP(84)  & 0.2     & 0.4  & the radius of the core Gaussian matter \\
  PARP(85)  & 0.33    & 0.9  & the probability that two gluons are produced in MPI with colors connecting to nearest neighbors\\
  PARP(86)  & 0.66    & 0.95  & the probability that two gluons are produced in MPI with the PARP(85) condition or as a closed loop \\
  PARP(89)  & 1000    & 1800  & reference energy scale for the turn-off $\pT$ \\
  PARP(90)  & 0.16    & 0.25  & energy dependence of the turn-off $\pT$ \\
  PARP(67)  & 1.0     & 4.0   & hard-scattering scale $\mu^2$ multiplied by this sets the maximum parton virtuality in initial-state radiation\\
\hline
  MSTP(51)  & 7       & 7    & CTEQ 5L PDF used. \\
  MSTP(91)  & 1       & 1    & Gaussian $k_T$ used. \\
  PARP(91)  & 1.0     & 1.0  & width of $k_T$ distribution. \\
  PARP(93)  & 5.0     & 5.0  & upper cutoff for $k_T$ dist. \\
\end{tabular} \end{ruledtabular}
\end{table}
We call a {\sc pythia} simulation with the tune-A setting `{\sc pythia} MPI',
although it has been adopted
as default values in the {\sc pythia} version 6.226 and later.

We use the output of the `{\sc pythia} default' and the `{\sc pythia} MPI' 
simulations to estimate the effect of the underlying event on our 
measurement.

The PHENIX experiment has developed its own {\sc geant}3-based detector simulator,
called the Phenix Integrated Simulation Application.
The absolute scale and the resolution of the EMCal energy and 
the tracking momentum have been tuned in the simulation 
using mass distributions of
$\pi^0$ ($2\gamma$), $\pi^\pm$, $K^\pm$ and $p^\pm$.

\subsubsection{Relation between $\pTreco$ and $\pTpy$}

The {\sc pythia}+{\sc geant} simulation was used to evaluate
the effect of the detector response and the underlying event on
the $\pTreco$ measurement.
The $\pT$ of a jet in {\sc pythia},
which is represented by $\pTpy$ in this paper,
should be defined so that 
it is comparable with the theoretical jet in order to evaluate
the relation between the NLO calculation and the measurement.
The event-by-event transition from the jet in {\sc pythia} ($\pTpy$)
to the reconstructed jet ($\pTreco$) is simulated to obtain
the statistical relation between them.

\begin{figure}[bthp] 
\includegraphics[width=1.0\linewidth]{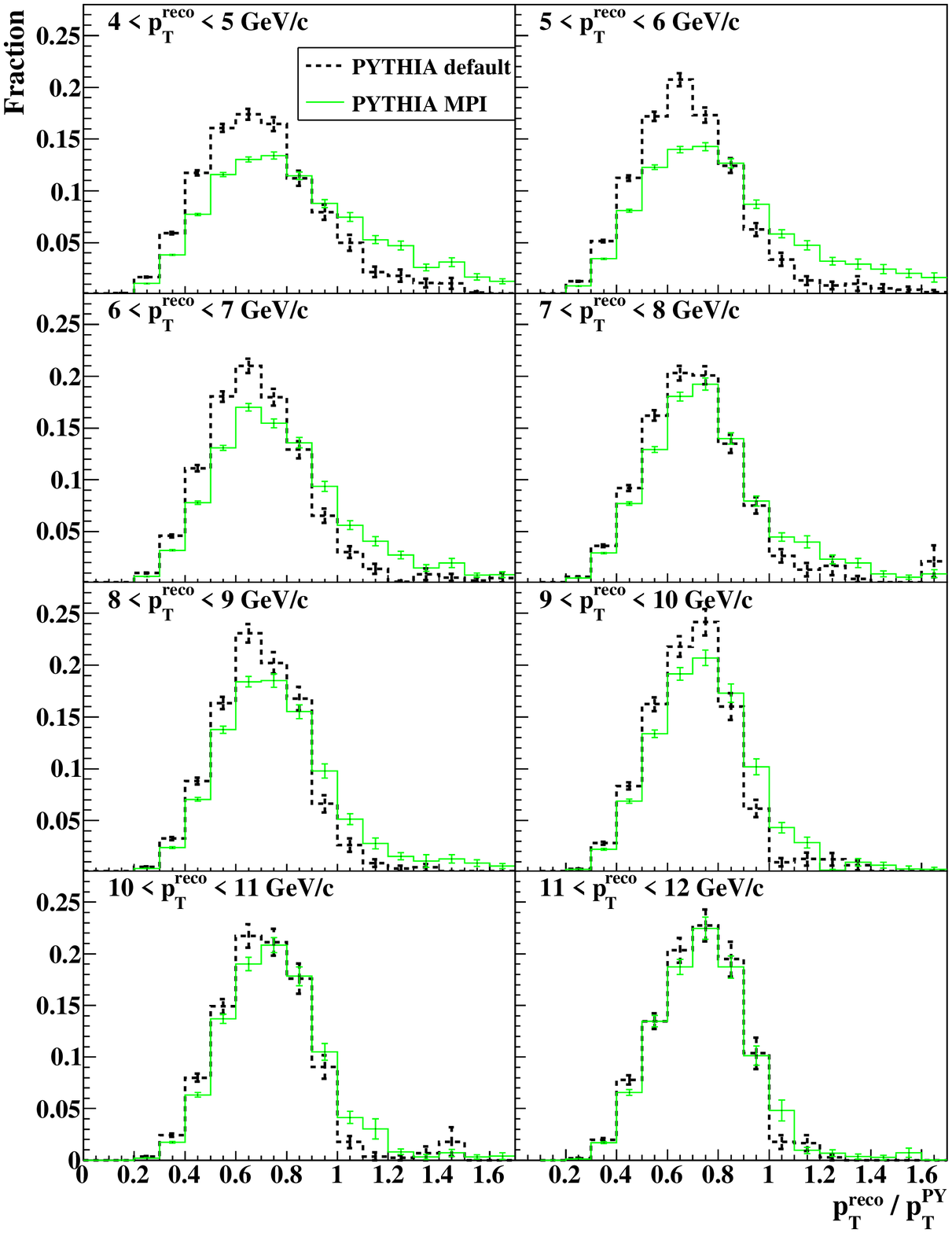}
\caption{ (color online) 
 Distributions of the ratio $\pTreco/\pTpy$ evaluated with 
(dashed black) {\sc pythia} default and (solid green)
{\sc pythia} MPI.
}
\label{fig:ratio}

\includegraphics[width=1.0\linewidth]{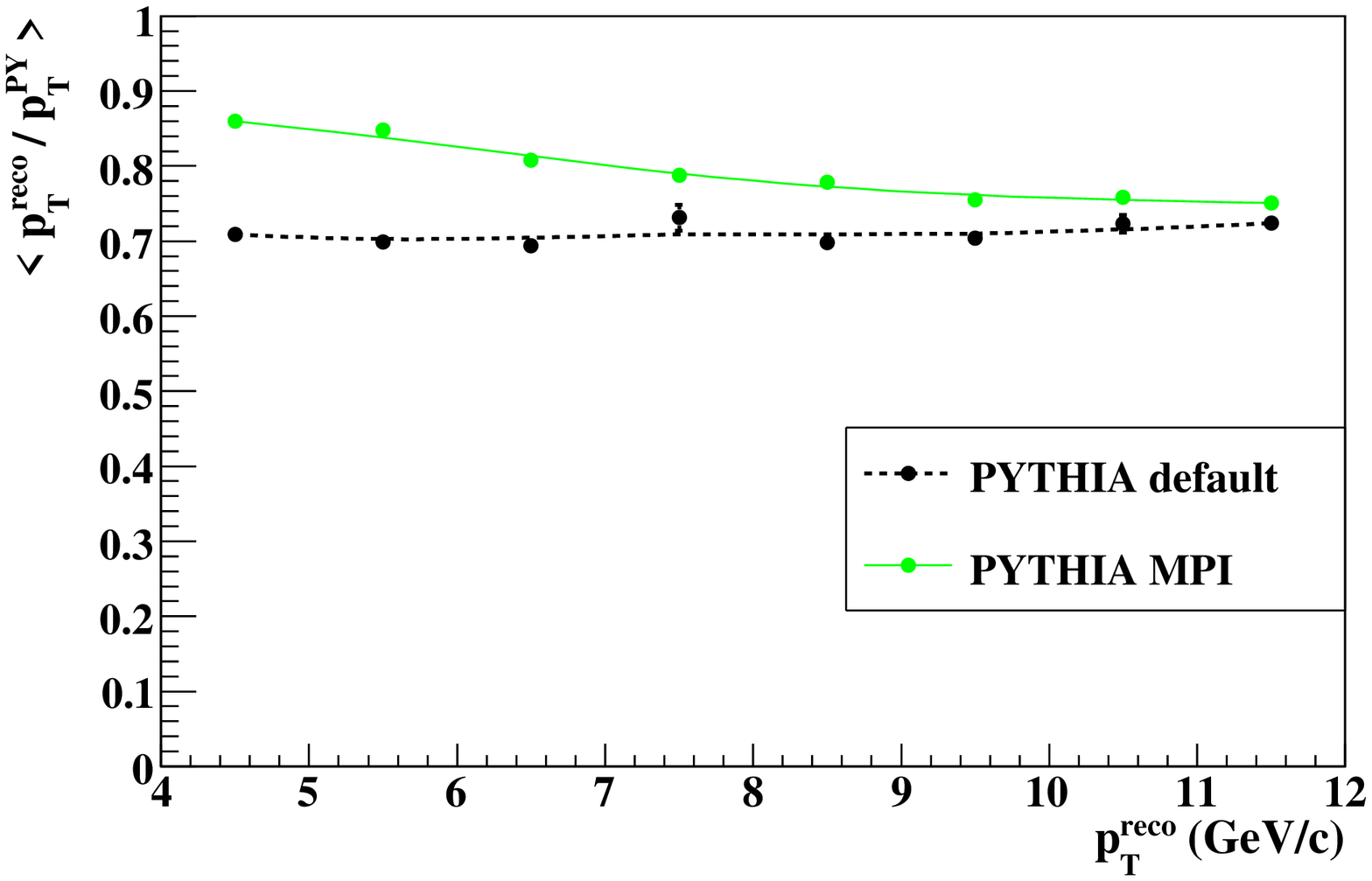}
\caption{ (color online) 
 The mean value of ratio $\pTreco/\pTpy$ as functions of $\pTreco$.
}
\label{fig:ratio_vs_pT_cone}
\end{figure}

A jet in {\sc pythia} is defined as a hard-scattered parton that 
has not undergone final-state parton splits,
namely particle number 7 or 8 in the {\sc pythia} event list.
A simulated reconstructed jet is associated with
one of the two partons by minimizing the angle
$\Delta R = \sqrt{\Delta\eta^2 + \Delta\phi^2}$.
Figure \ref{fig:ratio} shows
the ratio $\pTreco / \pTpy$ at each $\pTreco$ bin, and
Fig.~\ref{fig:ratio_vs_pT_cone} shows
the mean value of the ratios as a function of $\pTreco$.
The ratio of the {\sc pythia} MPI output is $\sim$80\% on average and is larger
than that of the {\sc pythia} default output due to the contribution from 
the underlying event.

The relation between reconstructed jets and jets in {\sc pythia} can be characterized
by multiple effects.
Some particles in a jet can leak from the cone
because of the limited acceptance, the small cone size
and the absence of a detector for neutral hadrons.
Some particles produced by the underlying event can be included
in the cone and contaminate $\pTreco$, and thus
the ratio $\pTreco/\pTpy$ can exceed one.
The $\pTpy$ of events that are in a $\pTreco$ bin is
distributed widely due to the finite $\pT$ resolution of the PHENIX
Central Arm.
Because a gluon jet is softer and broader than 
quark jet~\cite{Alexander:1991ce,Akers:1995ima},
the high-$\pT$ photon requirement has lower efficiency for gluon jets.
Therefore the ratio of $\pTreco$ to $\pTpy$ for gluon jets is smaller 
than quark jets on average.

\begin{figure}[bthp] 
\includegraphics[width=1.0\linewidth]{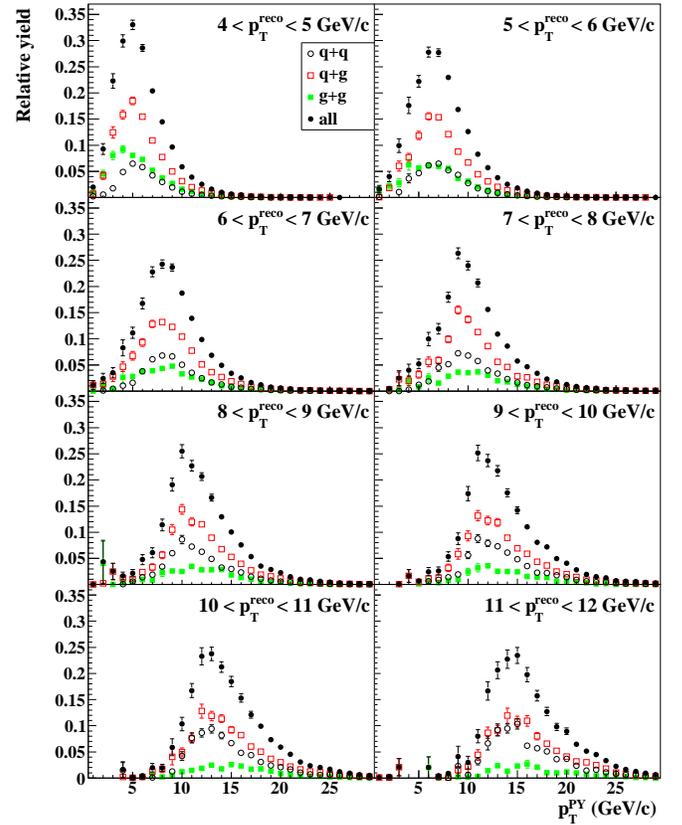}
\caption{ (color online) 
 The relative yields of $q+q$, $q+g$ and $g+g$ subprocesses 
 in the {\sc pythia}+{\sc geant} simulation.
 The results with all the subprocesses combined are also shown.
}
\label{fig:subproc_frac}
\end{figure}

\begin{figure}[bthp] 
\includegraphics[width=1.0\linewidth]{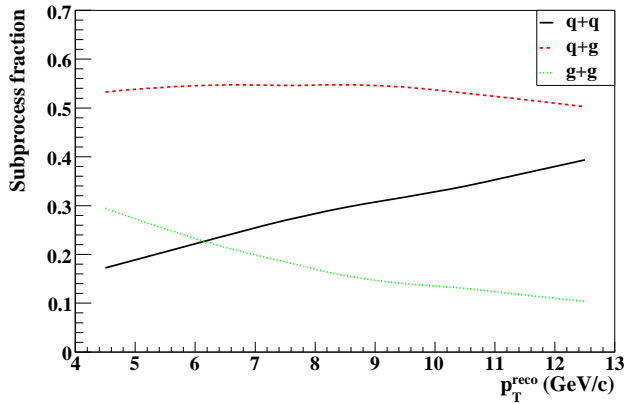}
\caption{ (color online) 
 Subprocess fractions of reconstructed jets as functions of $\pT$.
 It was evaluated with the {\sc pythia} MPI and {\sc geant} simulation.
 It should be noted that the gluon-quark reaction is the dominant reaction 
 in all the momentum region from 4 to 12 GeV/$c$.
}
\label{fig:subproc_frac_biased}
\end{figure}

Figure \ref{fig:subproc_frac} shows the relative yields of 
quark+quark ($q+q$), quark+gluon ($q+g$) and gluon+gluon ($g+g$)
subprocesses as a function of $\pTpy$ at each $\pTreco$ bin.
Figure \ref{fig:subproc_frac_biased} shows the fraction of $g+g$, $q+g$ and $q+q$
subprocesses as a function of $\pTreco$.
These were evaluated with the simulation.
As explained above,
the $gg$ subprocess is suppressed in this measurement.
The dominant subprocess is $q+g$ throughout the $\pTreco$ range.

\subsubsection{Relation between $\pTpy$ and $\pTnlo$}

\begin{figure}[bthp] 
\includegraphics[width=1.0\linewidth]{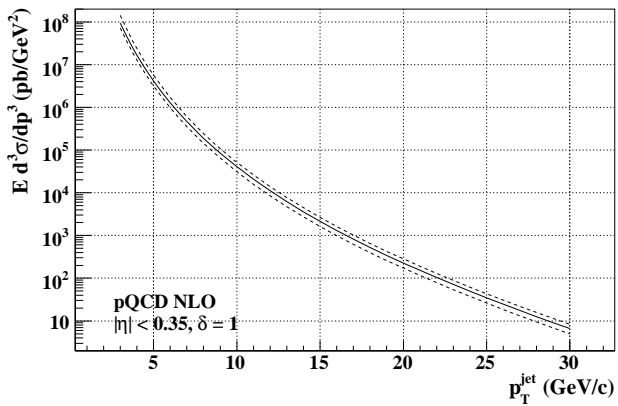}
\caption{
 Unpolarized jet cross section at a pseudorapidity $|\eta|<0.35$ with a
 cone half-aperture $\delta=1$.
 It was calculated at NLO under the SCA with three factorization scales,
 $\mu = \pT$ ({\bf solid line}), $2\pT$ ({\bf lower dashed line}) 
 and $\pT/2$ ({\bf upper dashed line}).
}
\label{fig:jet_xsec}
\end{figure}

The cross section and the $\ALL$ of inclusive jet production 
in $|\eta| < 0.35$ at $\sqrt{s} = 200$ GeV 
were calculated within the NLO pQCD framework 
with the CTEQ6M unpolarized PDF under the Small Cone
Approximation (SCA)~\cite{Aversa:1989xw,Jager:2004jh,Vogl_priv}.
We adopted a cone size of $\delta = 1.0$ for reasons 
that will be explained in a later section.
Figure \ref{fig:jet_xsec} shows the cross section calculated
with three factorization scales,  $\mu = \pT$, $2\pT$ and $\pT/2$
in NLO pQCD.

The $\pTnlo$ needs to be connected with $\pTpy$ in order to 
evaluate the relation between the NLO calculation and the measurement,
where the relation between $\pTpy$ and $\pTreco$
was obtained from the {\sc pythia}+{\sc geant} simulation.
We assume $\pTpy = \pTnlo$, and thus
the relation between the jet in {\sc pythia} and the measurement can be 
interpreted as the relation between the NLO calculation and the measurement.
However the definition of $\pTpy$ and $\pTnlo$ has a discrepancy,
and they become close to each other only as the cone half-aperture ($\delta$) 
in the theory becomes large.
Therefore we set $\delta$ to 1.0, which is the upper limit where
the SCA is applicable,
and evaluated the discrepancy between $\pTpy$ and $\pTnlo$ with 
$\delta = 1.0$ as described later.
Moreover,
the cone size of the jet in the NLO calculation needs to be larger than 
the acceptance of the PHENIX Central Arm 
so that one jet per central arm per event can be reconstructed
and connected with the jet in the NLO calculation.
This has been also satisfied with the use of $\delta = 1.0$.

Note that the cone size in theory and measurement
are different parameters and 
the difference is compensated for with the {\sc pythia} simulation; 
the former is related to the angle between two splitting partons and 
the latter is related to the angle between stable particles.

\subsubsection{Uncertainty due to difference in jet definitions}

The uncertainty due to the jet-definition difference between
the {\sc pythia} and NLO calculations with $\delta = 1.0$ 
has been evaluated using the difference between two jet definitions in {\sc pythia}.
One definition is the jet in {\sc pythia} defined above.
The other assumes a cluster of partons 
with a cone size of $\delta = 1.0$ in {\sc pythia},
where partons originating from the underlying event are excluded.
For the latter definition the jet $\pT$ is denoted $\pT^{\rm in\ cone}$.
Since $\pT^{\rm in\ cone}$ and $\pTnlo$ are defined similarly,
i.e.~both at the partonic level and with the same cone size $\delta$,
we assume that the scales of $\pT^{\rm in\ cone}$ and $\pTnlo$ are the same.
Then the difference between $\pT^{\rm in\ cone}$ and $\pTpy$, 
which can be evaluated using {\sc pythia},
is considered to be the difference between $\pTnlo$ and $\pTpy$.

\begin{figure}[bthp] 
\includegraphics[width=1.0\linewidth]{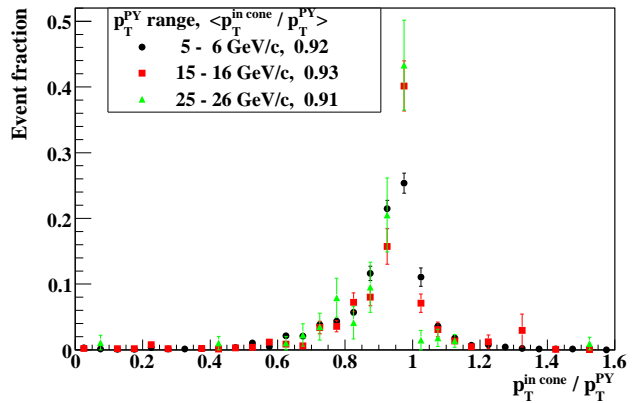}
\caption{ (color online) 
 Distributions of the fraction $p_T^{\rm in\ cone}/p_T^{\rm PY}$ 
 evaluated with a {\sc pythia} simulation at three typical $p_T^{\rm PY}$ bins.
}
\label{fig:h1_scale_diff}
\end{figure}

Figure \ref{fig:h1_scale_diff} shows distributions of 
the fraction $p_T^{\rm in\ cone}/p_T^{\rm PY}$ at three typical $p_T^{\rm PY}$ bins.
This indicates that 
the $p_T$ scales of the two jet definitions have a 10\% difference on average
in the $p_T$ range of these measurements.
Therefore the uncertainty due to the jet-definition difference between
{\sc pythia} and the NLO calculation with $\delta = 1.0$ 
has been assigned 10\% in $p_T$ scale.

\subsubsection{Reproducibility Check}

\begin{figure}[bthp] 
\includegraphics[width=1.0\linewidth]{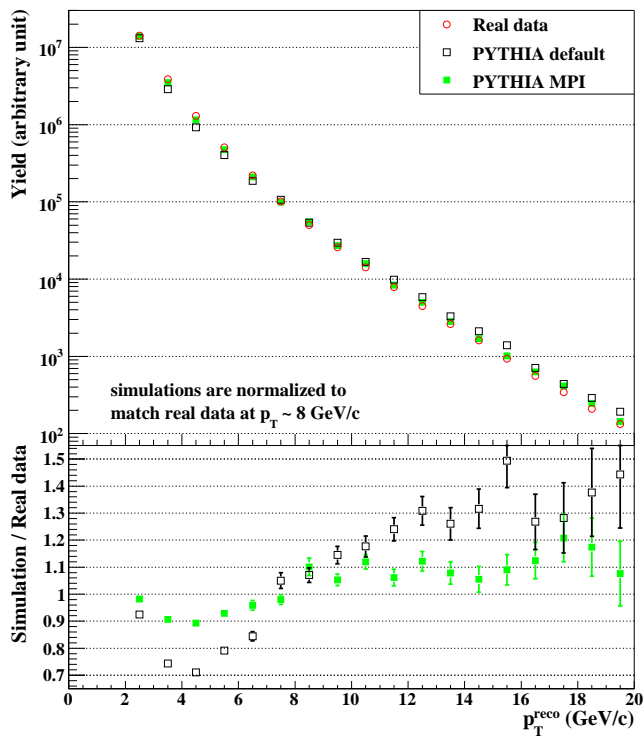}
\caption{ (color online) 
 Reconstructed-jet yields as a function of $\pTreco$.
 The red, black and green points correspond to the real data, the {\sc pythia}
 default output and the {\sc pythia} MPI output, respectively.
 The simulation outputs have been normalized so that they match the
 real data at $\pT = 8$ GeV/$c$.
 The ratio of the yields between the simulations and the real data
 is shown at bottom.
}
\label{fig:pT_cone_dist}
\end{figure}

Figure \ref{fig:pT_cone_dist} shows the distribution of $\pTreco$
measured with the clustering method described above.
The simulation outputs have been normalized so that they match
the real data at $\pT \sim 8$ GeV/$c$.
The slope of the {\sc pythia} MPI output agrees better with that of the real data,
where that of the {\sc pythia} default output is less steep.
The relative yield between the real data and the {\sc pythia} MPI output is
consistent within $\pm$10\% over five orders of magnitude.

\begin{figure}[bthp] 
\includegraphics[width=1.0\linewidth]{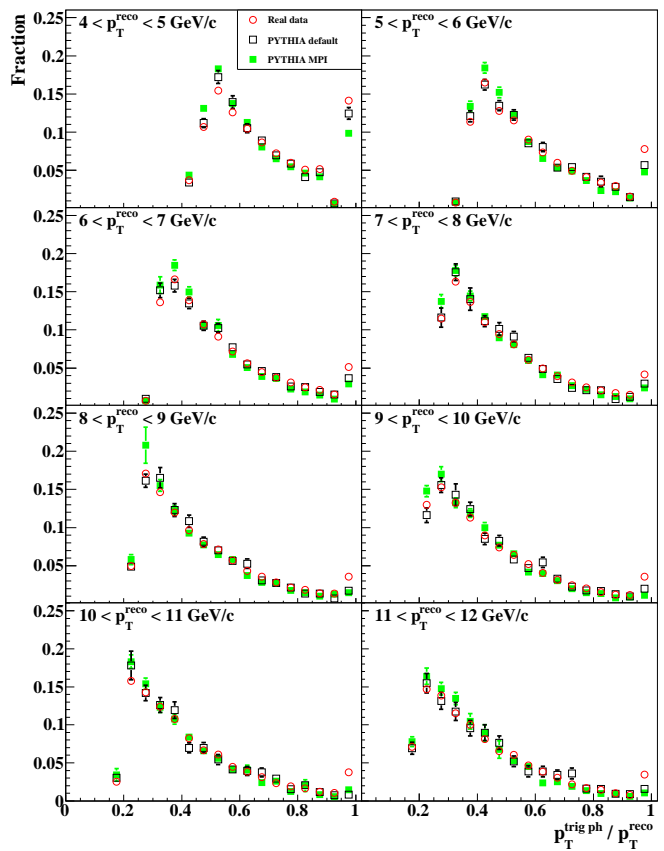}
\caption{ (color online) 
 The fraction of $\pT$ of the trigger photon in each $\pTreco$.
}
\label{fig:pT_lph_over_pT_cone}
\end{figure}

Figure \ref{fig:pT_lph_over_pT_cone} shows distributions of 
the fraction $\pT^{\rm trig\gamma} / \pTreco$, 
where $\pT^{\rm trig\gamma}$ is $\pT$ of the trigger photon.
The lower cutoff of the distributions is due to the minimum $\pT$ of
the trigger photon ($>2$ GeV/$c$).
The rightmost bin ($\pT^{\rm trig\gamma} / \pTreco \sim 1$) contains
events in which only a trigger photon exists.
Such events can occur by the limited acceptance, by the EMCal masked area
(particles except a trigger photon in jet are not detected), by EMCal
noise or by direct photon events.
The difference between the real data and the simulation outputs 
in the rightmost bin may indicate that these effects are not completely
reproduced by the simulation,
but the difference is small ($<5\%$) and negligible
in comparison with other uncertainties.

\section{Results and discussions}

\subsection{Event structure}

\subsubsection{Multiplicity}

\begin{figure}[bthp] 
\includegraphics[width=1.0\linewidth]{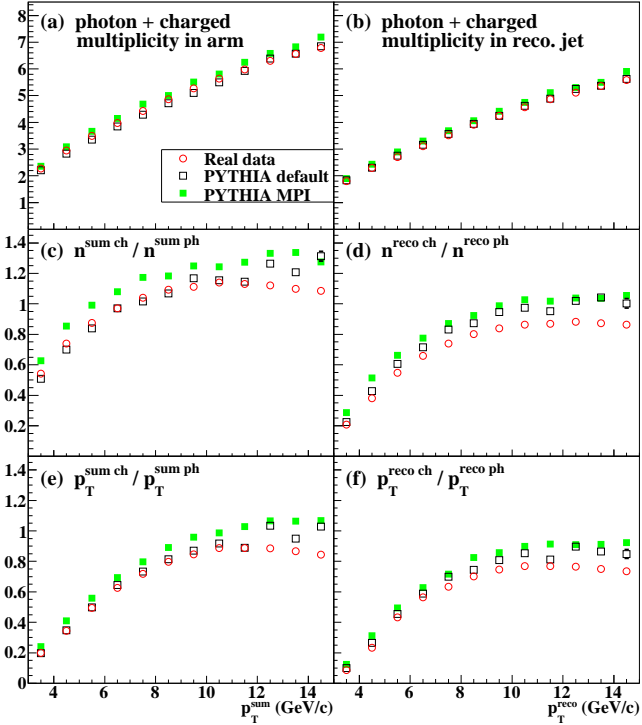}
\caption{ (color online) 
 {\bf (a)}: Mean multiplicity in the Central Arm vs $\pTsum$.
 {\bf (b)}: Mean multiplicity in the cluster vs $\pTreco$.
 {\bf (c)}: The ratio of charged multiplicity to photon multiplicity in the Central Arm.
 {\bf (d)}: Same as (c) but in the cluster.
 {\bf (e)}: The ratio of charged $\pT$ to photon $\pT$ in the Central Arm.
 {\bf (f)}: Same as (e) but in the cluster.
}
\label{fig:multiplicity}
\end{figure}

Multiplicity is defined as the number of particles which satisfy the
experimental cuts in one event.
Figure \ref{fig:multiplicity}(a) and (b) show 
the mean value of multiplicity in
the Central Arm vs $\pTsum$ and in the cluster vs $\pTreco$.
The multiplicities in the arm and in the cluster of the simulation outputs agree, 
on the whole, with that of the real data.
The {\sc pythia} MPI output is larger than the {\sc pythia} default output as
expected, and the real data are closer to the {\sc pythia} default output.
On the other hand,
the $\pTreco$ distributions (Fig.~\ref{fig:pT_cone_dist}) shows better
agreement between the real data and the {\sc pythia} MPI output.
This indicates that the {\sc pythia} MPI reproduces the sum of $\pT$ of particles
well, which is less sensitive to particle fragmentation process, while
it does not reproduce the particle multiplicity very well.
The reproducibility of the summed $\pT$ is checked in measurements described later.

Figure \ref{fig:multiplicity}(c) and (d) show
the ratio of charged-particle multiplicity to photon multiplicity
in the Central Arm and in the cluster.
The real data lies below the {\sc pythia} default and MPI
results for both multiplicities.
This indicates that the effect of the underlying event in the ratios is
small, and the difference between the real data and the
{\sc pythia} results is mainly caused by the imbalance between 
photons and charged particles in jet.
Figure \ref{fig:multiplicity}(e) and (f) show
the ratio of the sum of charged-particle $\pT$ to the sum of photon $\pT$.
These have the same tendency as the multiplicity ratios.

\subsubsection{Transverse momentum density}

The $\pT$ density, ${\cal D}_{P_T} (\Delta \phi)$, is defined as
\begin{equation}
{\cal D}_{\pT} (\Delta \phi) \equiv
  \left< \frac{1}{\delta \phi} 
     \sum_{i {\rm  in } [\Delta \phi,\ \Delta \phi + \delta \phi]} 
     p_{Ti} \right>_{\rm event},
\end{equation}
where $\Delta \phi$ is $\phi$ angle with respect to the direction of a
trigger photon in event, $\delta \phi$ is an area width
in $\phi$ direction, and $p_{Ti}$ is
transverse momentum of $i$-th particle in event.
The $\pT$ density means the area-normalized total
transverse momentum in an area of $\delta \phi \times \delta \eta$ at
a distance $\Delta \phi$ from trigger photon,
where $\delta\eta$ is the width of the Central Arm acceptance.

We name the region at $\Delta \phi \lesssim 0.7$ rad the `toward' region and the
region at $\Delta \phi \gtrsim 0.7$ rad the `transverse' region.
Since particles from a jet are concentrated along the jet direction, the
${\cal D}_{\pT}$ in the transverse region is sensitive to the
underlying event.

\begin{figure}[hb] 
\includegraphics[width=0.4\linewidth]{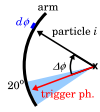}
\caption{ (color online) 
 Measurement condition of the $\pT$ density.
 The arc and the $\times$ mark represent 
 one Central Arm and the collision point in the beam view.
}
\label{fig:explain_pT_density}
\end{figure}

As illustrated in Fig.~\ref{fig:explain_pT_density},
to avoid the effect of the PHENIX Central Arm acceptance in the
calculation of ${\cal D}_{\pT}$, we limited the $\phi$ direction of the trigger
photons to less than 20$\Deg$ from one edge of the PHENIX Central Arms,
and we did not use photons and charged particles which were in the $\phi$
area between the trigger photon and the near edge.
With this method the ${\cal D}_{\pT}$ distribution is not affected
by the finite acceptance of the PHENIX Central Arms up to 70$\Deg$
($\sim$ 1.2 rad).
\begin{figure}[th] 
\includegraphics[width=1.0\linewidth]{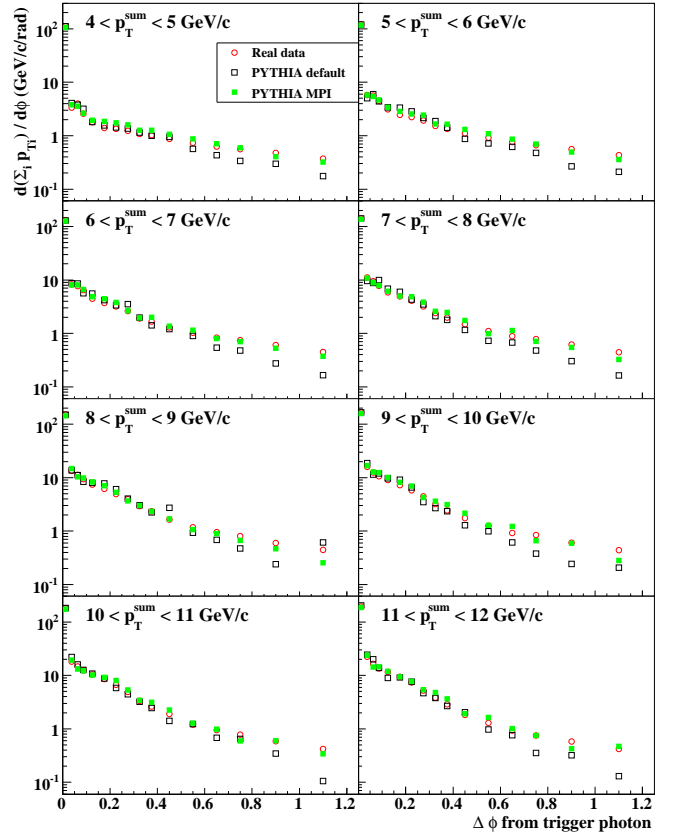}
\caption{ (color online) 
 $\pT$ density, 
 ${\cal D}_{P_T} = d\Sigma_i p_{Ti}/d\phi$ (GeV/$c$/rad), 
 in each $\pTsum$ bin.
 Trigger photons are included in the leftmost points.
}
\label{fig:pT_density}
\end{figure}

Figure \ref{fig:pT_density} shows the 
${\cal D}_{\pT}$ distributions for each $\pTsum$ range.
In the ``toward'' region, 
the simulation outputs agree well with the real data.
It shows that the shape of jets produced by the simulation is consistent
with the real data.
In the ``transverse'' region, the {\sc pythia} default output is generally
smaller than the real data.
This is an indication that the {\sc pythia} default does not contain
sufficient total $\pT$ of soft particles from the underlying event.
The {\sc pythia} MPI output agrees with the real data well.

\subsubsection{Thrust distribution in PHENIX Central Arm}

We evaluated the thrust variable defined in the CERN-ISR era with particles in
one PHENIX Central Arm ($\Delta \eta = 0.7$, $\Delta \phi = 90^{\rm o}$):
\begin{equation}
T_{PH} \equiv \max_{\Vec{u}}
      \frac{\sum_i |\Vec{p}_i \cdot \Vec{u}|}{\sum_i |\Vec{p}_i|}
 = \frac{\sum_i |\Vec{p}_i \cdot \hat{\Vec{p}}|}{\sum_i |\Vec{p}_i|}
    \label{eq:thrust} 
\end{equation}
\begin{equation}
\hat{\Vec{p}} = \frac{\sum_i \Vec{p}_i}{|\sum_i \Vec{p}_i|},
\end{equation}
where $\Vec{u}$ is a unit vector which is called the thrust axis and is
directed to maximize $T$, and $\Vec{p}_i$ is a momentum of each
particle in one arm.
If only particles in a half sphere in an event are used,
$T_{PH}$ can be written as the right-side formula in Eq.~\ref{eq:thrust}.

The distribution of $T_{PH}$ of isotropic events in the PHENIX Central
Arm acceptance for each $\pTsum$ bin was simulated with the
following method.
First, 
the cross section of inclusive particle production is assumed to be
proportional to $\exp(- 6\ \pT ({\rm GeV}/c))$ and is independent
of $\eta$ and $\phi$.
Second, 
the same cuts as the experimental conditions are applied numerically: 
the geometrical acceptance ($|\eta| < 0.35$, $\Delta \phi = 90^{\rm o}$),
the momentum limit ($\pT > 0.4 {\rm  GeV}/c$), and one high-$\pT$
particle ($\pT > 2.0 {\rm  GeV}/c$).
Third, 
the distribution of $T_{PH}$ of isotropic events was calculated for
each number of particles in one event 
($f_n(T)$ for $n = 1, 2, 3, \dots$).
The $T_{PH}$ distribution of $n=2$ events is particularly steep.
Thus we applied a cut of $n \ge 3$ in the $T_{PH}$ measurement.
The $f_T$ is evaluated as the sum of $f_n(T)$'s weighted by the probability
($\epsilon_n$) that the number of particles per event is $n$:
\begin{equation}
f(T) = \sum_n \epsilon_n f_n(T) \ \ ,\ \ \ 
\epsilon_n = \frac{N_{\rm evt}^n}{N_{\rm evt}}, 
\label{eq:thrust_iso}
\end{equation}
where $\epsilon_n$ was derived from the real data.

\begin{figure}[bthp] 
\includegraphics[width=1.0\linewidth]{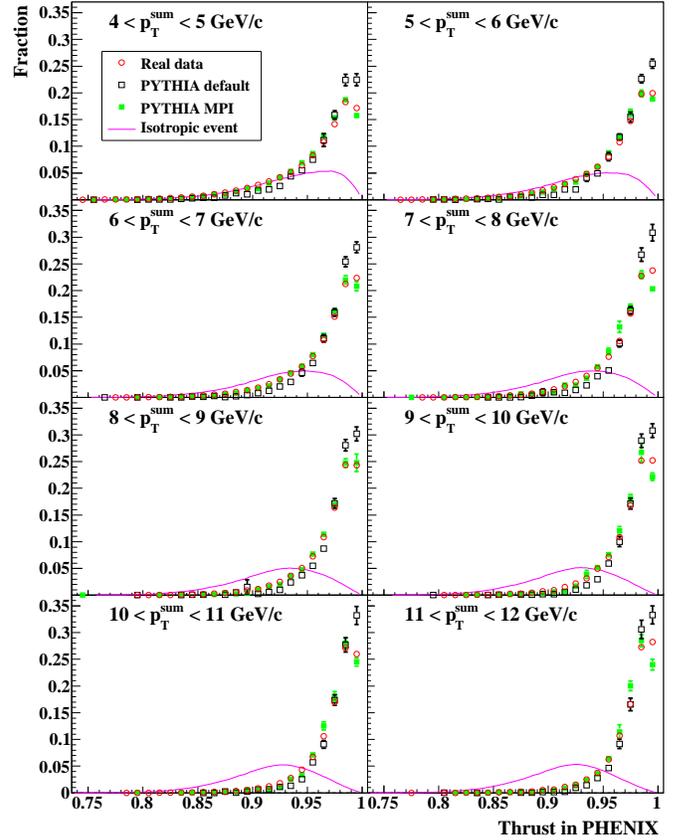}
\caption{ (color online) 
 $T_{PH}$ distribution in each $\pTsum$ bin.
 All distributions have been normalized 
 so that their areas were equal to one another.
 The purple lines are the distributions of isotropic events in the
 acceptance of the PHENIX Central Arms, which are evaluated with
 Eq.~\ref{eq:thrust_iso}.
}
\label{fig:thrust}
\end{figure}

Figure \ref{fig:thrust} shows the $T_{PH}$
distribution in each $\pTsum$ range.
The {\sc pythia} MPI output agrees with the real data well.
The {\sc pythia} default has a steeper slope, which indicates
that the number of particles in the vicinity of jets 
in the {\sc pythia} default is insufficient.
In the real data, the {\sc pythia} default output and the {\sc pythia} MPI
output, the $T_{PH}$ distribution becomes sharper as $\pTsum$ increases. 
This is due to the fact that the transverse
momentum ($j_T$) of a jet is independent of its longitudinal momentum 
and is almost constant.

If the real data includes a contribution from non-jet (isotropic)
events,
the $T_{PH}$ distribution of the real data is a mixture of the
distribution of the simulation output and the distribution of the
isotropic case.
The contribution from non-jet events can be judged to be negligible
because the {\sc pythia} MPI output reproduces the data 
even though it does not have isotropic events.

\subsection{Jet production rate}

\subsubsection{Evaluation method (measurement)}

The jet production rate ${\cal Y}$, 
namely the yield of reconstructed jets per unit
luminosity, is defined with measured quantities as
\begin{equation}
{\cal Y}^{i} \equiv \frac{N_{reco}^{i}}{L \cdot f_{\rm MB} \cdot f_{ph}},
\end{equation}
where 
$L$ is the integrated luminosity;
$f_{\rm MB}$ and $f_{ph}$ are the efficiencies of the MB trigger 
(see Sec.~\ref{sec:trigger}) and the high-$\pT$ photon trigger, respectively;
$N_{reco}^{i}$ is the reconstructed-jet yield in a $i$-th $\pTreco$ bin.
The high-$\pT$ photon trigger efficiency $f_{ph}$ was estimated to be
$0.92 \pm 0.02$, where the inefficiency is caused by the 
10\% of the EMCal acceptance where the trigger was 
disabled due to electronics noise.
The inefficiency is slightly smaller than the disabled acceptance because
a particle cluter can contain multiple high-$\pT$ photons.

\subsubsection{Evaluation method (prediction)}

On the other hand,
the variable ${\cal Y}$ is expressed 
with theoretical and simulation quantities as
\begin{equation}
{\cal Y}^{i} \equiv 
\sum_{j} f^{ij} \cdot \epsilon_{trig+acc}^{j} \cdot {\cal Y}_{theo}^{j},
\end{equation}
where
the label $i$ and $j$ are the indices of $\pTreco$ and $\pTnlo$ bins,
respectively.
The ${\cal Y}_{theo}^{j}$ is a jet production rate within 
$|\eta| < 0.35$ in a $j$-th $\pTnlo$ bin, which is theoretically calculated.
The $\epsilon_{trig+acc}^{j}$ is a high-$\pT$-photon trigger efficiency
and acceptance correction, which is evaluated with the {\sc pythia}+{\sc geant} simulation.
The $\epsilon_{trig+acc}^{j} \cdot {\cal Y}_{theo}^{j}$ is a yield of jets
that include a high-$\pT$ photon within $|\eta| < 0.35$.
The $f^{ij}$ is the probability that a jet within a
$j$-th $\pTnlo$ bin is detected as a reconstructed jet 
within a $i$-th $\pTreco$ bin.
This method uses the relative $\pTreco$ distribution in each $\pTnlo$
bin and thus the slope of the $\pTpy$ distribution in the simulation
does not affect the result of ${\cal Y}^{i}$.

\begin{figure}[bthp] 
\includegraphics[width=1.0\linewidth]{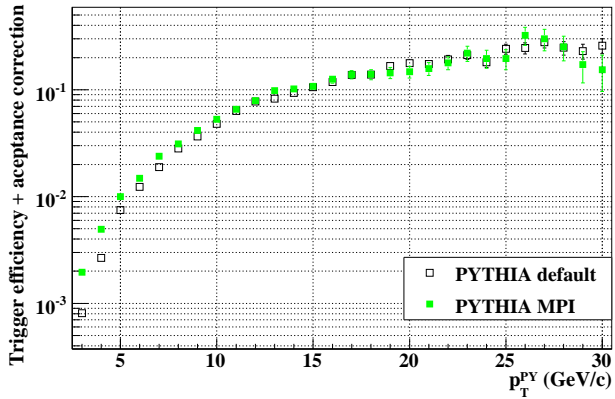}
\caption{ (color online) 
 The correction factor $\epsilon_{trig+acc}^{j}$ for high-$p_T$ photon
 trigger efficiency and acceptance effect.
 The {\sc pythia} default ({\bf black}) and the {\sc pythia} MPI setting 
 ({\bf green}) were used.
}
\label{fig:trig_and_acce}
\end{figure}

The correction factor $\epsilon_{trig+acc}^{j}$ is a fraction,
whose numerator is the number of events in which 
at least one photon with $\pT > 2$ GeV/$c$ is detected,
and whose denominator is the number of events in which
jets are in $|\eta| < 0.35$.
The condition ``$\pT > 2$ GeV/$c$'' in the numerator corrects
a high-$\pT$ photon efficency, i.e.~the probability that 
a high-$\pT$ photon in jets must be detected with the EMCal.
The condition ``$|\eta| < 0.35$'' in the denominator 
and the absence of it in the numerator corrects an acceptance for jets,
i.e.~the fact that a part of reconstructed-jets does 
originate from jets with $|\eta| > 0.35$.
Figure \ref{fig:trig_and_acce} shows
$\epsilon_{trig+acc}^{j}$ as a function of $\pTnlo$ estimated with the
{\sc pythia} default and MPI simulations.

To estimate a systematic error related to the simulation reproducibility of 
high-$\pT$ photon,
we evaluated, in both the real data and the simulations,
the ratio ($r$) of the reconstructed-jet yields in the high-$\pT$ photon triggered
sample to that in the MB triggered sample.
The $r$ of the {\sc pythia} MPI output is 5\% at $\pTreco = 4$ GeV/$c$
and 50\% at $\pTreco = 12$ GeV/$c$,
and is consistent with that of the real data within $\pm$10\%.
Therefore a 10\% error was assigned to the jet production rate
calculated with the {\sc pythia} MPI simulation.
The $r$ of the {\sc pythia} default output is smaller by 20-30\% than that of 
the real data.

\subsubsection{Result}

\begin{figure}[bthp] 
\includegraphics[width=1.0\linewidth]{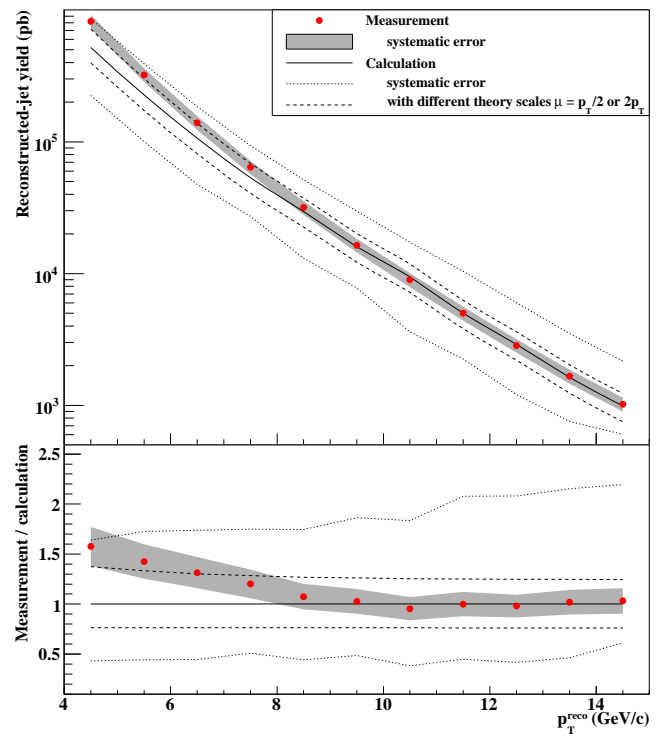}
\caption{ (color online) (top) Reconstructed jet yield 
and (bottom) the ratio of the real data to the calculations.
(red points) Real data 
with the (gray band) experimental systematic error.
(black curves) {\sc pythia} MPI calculation with theory 
factorization scales of 
(solid curve) $\pT$, (upper dashed curve) $\pT/2$, and  
(lower dashed curve) $2\pT$. 
(dotted curves) Variation caused by 10\% $\pT$ scale uncertainty
around the calculation.   Statistical uncertainties are smaller than 
the size of the points.
}
\label{fig:event_rate}
\end{figure}

Figure \ref{fig:event_rate} shows the jet production rate.
The main systematic errors are listed in Tab.~\ref{tab:sys_error}.
The main uncertainties of the measurement are
the BBC cross section and the EMCal energy scale.
These errors are fully correlated bin-to-bin.
The error on the EMCal energy scale includes both the change of $\pT$ of
individual photons and the change of the threshold of the high-$\pT$
photon requirement.
In comparing the measurement and the calculation,
the 10\% $\pT$ scale uncertainty of the jet definitions
in the {\sc pythia} simulation and the NLO pQCD theory
makes a 30\% error at low $\pT$ or 70\% at high $\pT$, 
and is the largest source.
The uncertainty of the renormalization and factorization scales in the
NLO jet production cross section makes a 30\% error.
The calculation with {\sc pythia} MPI agrees with the measurement within
errors over the measured range $4 < \pTreco < 15$ GeV/$c$.

\begin{table}[bthp] 
\caption{Main systematic errors of the jet production rate. }
\label{tab:sys_error}
\begin{ruledtabular} \begin{tabular}{lcc}
 Source                    & Size  &  Size on rate \\
 \hline
 \multicolumn{3}{c}{Measurement} \\
 Luminosity                  & 9.7\%     & 9.7\%        \\
 EMCal energy scale          & 1.5\%     & 7-6\%   \\
 Tracking momentum scale     & 1.5\%     & 0-3\%   \\
 \hline
 \multicolumn{3}{c}{Calculation} \\
 Jet definition                  & 10\% in $\pT$ & 30-70\% \\
 Jet shape \& underlying event   & --            & 50-20\% \\
 High-$\pT$ photon fragmentation & --            & 10\%     \\
 Simulation statistics           & --            & 2-5\%    \\
\end{tabular} \end{ruledtabular}
\end{table}

The result with {\sc pythia} default is smaller than the result with
{\sc pythia} MPI by 50\% at $\pTreco$ = 4 GeV/$c$, 
by 35\% at $\pTreco$ = 9 GeV/$c$ and by 20\% at $\pTreco$ = 14 GeV/$c$.
It can be fully explained by the difference visible in
Fig.~\ref{fig:trig_and_acce} between {\sc pythia} default and {\sc pythia} MPI.
According to the comparisons of the event structure,
{\sc pythia} MPI reproduces the spatial distribution of particle momenta
in one event much better than the {\sc pythia} default.
Therefore, for the jet production rate evaluated with {\sc pythia} MPI
simulation, the error due to possible insufficient tunings of {\sc pythia}
MPI should be smaller than the difference of the jet production rate
between the {\sc pythia} MPI simulation and the {\sc pythia} default simulation.

\subsection{Double helicity asymmetry $A_{LL}$}

\subsubsection{Evaluation method (measurement)}

$\ALL$ is expressed with measured quantities as
\begin{equation}
\ALL = \frac{1}{|P_B| |P_Y|} 
\frac{(N_{++} + N_{--}) - R (N_{+-} + N_{-+})}
     {(N_{++} + N_{--}) + R (N_{+-} + N_{-+})}
\end{equation}
\begin{equation}
R \equiv \frac{L_{++} + L_{--}}{L_{+-} + L_{-+}},
\end{equation}
where
$N_{++}$ etc.~are reconstructed-jet yields with 
colliding proton beams having the same ($++$ or $--$) and opposite ($+-$
or $-+$) helicity;
$P_B$ and $P_Y$ are the beam polarizations;
$R$ is the relative luminosity, i.e.~the ratio of the luminosity with 
the same helicity ($L_{++} + L_{--}$) to that 
with the opposite helicity ($L_{+-} + L_{-+}$).
$\ALL$ is measured fill-by-fill and the results are fit to a
constant, because the beam polarization and the relative luminosity are
evaluated fill-by-fill to decrease systematic errors.
The average fill length was about five hours.
The integrated luminosity used was 2.1 pb$^{-1}$.
It is 0.1 pb$^{-1}$ less than the statistics used in 
the production rate measurement because the data with 
bad conditions on the beam polarization were discarded.

The relative luminosity at PHENIX was evaluated with the
MB trigger counts ($N_{\rm MB}^{++}$ and $N_{\rm MB}^{+-}$) as
$R = N_{\rm MB}^{++} / N_{\rm MB}^{+-}$.
A possible spin dependence of MB-triggered data causes an
uncertainty on the relative luminosity.
The error has been checked by comparing the relative luminosity with
another relative luminosity defined with the ZDCLL1 trigger counts.
The ZDCLL1 trigger is fired when both the north ZDC and the south ZDC
have a hit and the reconstructed $z$-vertex is 
within 30 cm of the collision point.
 
The beam polarizations were measured with 
the pC and H-jet polarimeters~\cite{Tojo:2002wb,Okada:2005gu} 
at the 12 o\'clock interaction point on the RHIC ring.
One of the colliding beam rotating clockwise is called ``blue beam'',
and the other rotating counterclockwise ``yellow beam''.
The luminosity-weighted-average polarizations are 
50.3\% for the blue beam and 48.5\% for the yellow beam.
The sum of statistical and systematic errors on
$\langle P_B \rangle \langle P_Y \rangle$ is 9.4\%.

\subsubsection{Evaluation method (prediction)}

\begin{figure}[bthp] 
\includegraphics[width=1.0\linewidth]{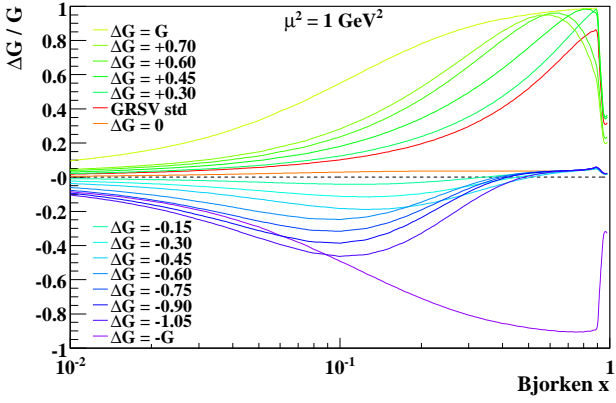}
\caption{ (color online)
 Assumed gluon distribution functions at $\mu^2$ = 1 GeV$^2$.
 The integral $\int_0^1 dx\Delta G(x)$ of each distribution 
 at the initial scale $\mu^2$ = 0.4 GeV$^2$ is, from bottom to top at $x=0.15$,
 -1.24 ($\Delta G = -G$), -1.05, -0.90, -0.75, -0.60, -0.45, -0.30, -0.15,
 0 ($\Delta G = 0$), 0.24\ (GRSV-std), 0.30, 
 0.45, 0.60, 0.70 and 1.24 ($\Delta G = G$).
}
\label{fig:dg_over_g_various}
\end{figure}

Polarized/unpolarized cross sections of jet production for every
subprocess ($q+q$, $q+g$ and $g+g$) were calculated at NLO based on the SCA
with a cone size of $\delta = 1.0$.
The polarized cross sections were calculated using various $\Delta G(x)$
in order to compare the measured $\ALL$ with various predicted $\ALL$'s
and find the most-probable $\Delta G(x)$.
Figure \ref{fig:dg_over_g_various} shows the distributions of 
the $\Delta G(x)$ used, and the integrated values are
\begin{align}
&\int_0^1 dx \Delta G(x, \mu^2 = 0.4 {\rm  GeV}^2) \nonumber \\
&= 
\begin{cases}
  -1.24\ (\Delta G = -G), \\
  -1.05, \\
  -0.90, \\
  -0.75, \\
  -0.60, \\
  -0.45, \\
  -0.30, \\
  -0.15,
\end{cases}
\begin{cases}
  0\ (\Delta G = 0), \\
  0.24\ ({\rm GRSV-std}), \\
  0.30, \\
  0.45,\\
  0.60, \\
  0.70, \\
  1.24\ (\Delta G = G) 
\end{cases}
\end{align}
Each $\Delta G(x)$ (except the GRSV-std, the $\Delta G = G$ input, the
$\Delta G = 0$ input and the $\Delta G = -G$ input) have been obtained
by refitting the GRSV parameters to the DIS data which were 
used in the original GRSV analysis~\cite{Gluck:2000dy}.
It is noted that
the DIS data used in GRSV are the data up to the year 2000 and thus
are much less than that used in the updated analysis, 
DSSV~\cite{deFlorian:2008mr}, for example.
The polarized PDF in the GRSV parameterization is of the form:
\begin{equation}
\Delta f(x, {\mu_0}^2) = 
    N_f x^{\alpha_f} (1-x)^{\beta_f} f(x, {\mu_0}^2)_{\rm GRV}, 
\label{eq:grsv_param}
\end{equation}
where 
$f$ is $u$, $d$, $\bar{q}$ or $G$;
${\mu_0}^2 = 0.4$ GeV$^2$ is the initial scale at which the
functional forms are defined as above;
$f(x, {\mu_0}^2)_{\rm GRV}$ is the unpolarized PDF of the
GRV98 analysis~\cite{Gluck:1998xa};
$N_f$, $\alpha_f$ and $\beta_f$ are free parameters.
In the refit of the DIS data,
the integral value of $\Delta G(x)$ from $x = 0$ to $1$
was fixed to its particular value
listed above, and the shape of $\Delta G(x)$ and the quark-related
parameters were made free.
The $\chi^2$ of the refitting to the DIS data is 
170 for the 209 data points~\cite{Gluck:2000dy}
when the integral of $\Delta G$ is 0 at the initial $\mu^2$, for example.
In the remainder of this paper
we concentrate on investigating the $\chi^2$ of the six data points
of the reconstructed-jet $\ALL$.

The various $\Delta G(x)$ above were evolved up to a scale $\mu$ of every
event in the $\ALL$ calculation.
The $\ALL$ of every subprocess ($\ALL^{q+q}$, $\ALL^{q+g}$ and $\ALL^{g+g}$) 
can be derived as functions of $\pTnlo$ from the unpolarized and
polarized cross sections.
The {\sc pythia}+{\sc geant} simulation produces the relative yields of every
subprocess ($n^{q+q}(\pTnlo, \pTreco)$, $n^{q+g}(\pTnlo, \pTreco)$ and 
$n^{g+g}(\pTnlo, \pTreco)$),
as shown in Fig.~\ref{fig:subproc_frac}.
$\ALLcone(\pTreco)$ is calculated as a mean of $\ALL^{q+q}$, $\ALL^{q+g}$ and
$\ALL^{g+g}$ weighted by the fractions of events:
\begin{align}
\lefteqn{\ALLcone(\pTreco)} \nonumber \\
& = \frac{\displaystyle \int d\pTnlo \sum_{i{\rm sub}} 
n^{i{\rm sub}}(\pTnlo, \pTreco) 
\cdot \ALL^{i{\rm sub}}(\pTnlo) }
     {\displaystyle \int d\pTnlo \sum_{i{\rm sub}} n^{i{\rm sub}}(\pTnlo, 
\pTreco) }, 
\end{align}
where $i$sub is $q+q$, $q+g$ and $g+g$.
As an estimation of systematic errors, 
the slope of jet yields and the fraction of subprocesses were compared
between the theory calculation and the {\sc pythia} simulation.
Note that both the slope and the fraction that we compared
have not been biased by the high-$\pT$ photon and the small cone, 
since the theory calculation cannot provide biased values.
The variations of $\ALLcone$ caused by both the slope difference
and the fraction difference are negligible in comparison with other errors.

\subsubsection{Result}

\begin{figure}[bthp] 
\includegraphics[width=1.0\linewidth]{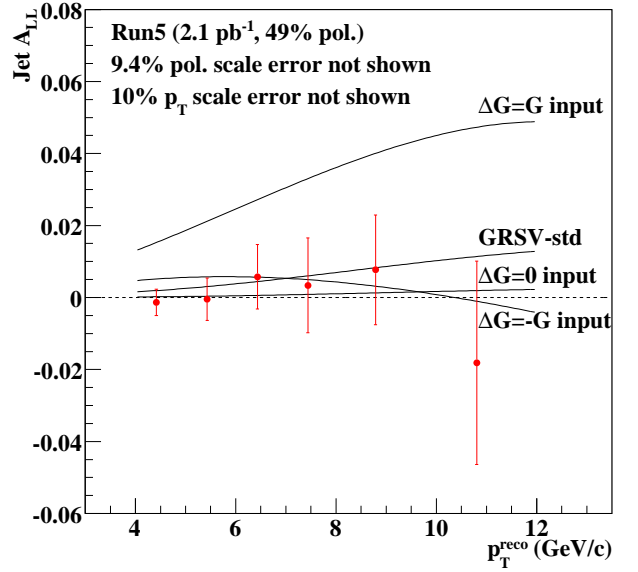}
\caption{ (color online) 
 Reconstructed-jet $\ALL$ as a function of $\pTreco$.
 (red points) Measurement with statistical error bars.
 (black lines) Calculation based on four $\Delta G(x)$
 functions and the {\sc pythia} MPI + {\sc geant} simulation.
}
\label{fig:A_LL_theo}
\end{figure}

Figure \ref{fig:A_LL_theo} shows measured $\ALLcone$ and four
prediction curves. 
Table \ref{tab:A_LL} shows the values of measured $\ALLcone$.
The measured $\ALL$ is consistent with zero,
as the $\chi^2$/n.d.f.~between the data points and 
zero asymmetry ($\ALL = 0$) is 1.3/6.
The systematic error of the relative luminosity is much smaller than the
statistical error on $\ALL$ and is negligible.
On the prediction curves the systematic error related to the fractions
of subprocesses are smaller than the 10\% $\pT$ scale uncertainty
by roughly an order of magnitude.
Therefore it is not included in this plot.

\begin{table}[bthp] 
\caption{ Measured reconstructed-jet $\ALL$. }
\label{tab:A_LL}
\begin{ruledtabular} \begin{tabular}{ccc}
$\pTreco$ range and mean (GeV/$c$) & $\ALL$ & stat~error \\
\hline
 4-5,   4.42 &  -0.0014  & 0.0037 \\
 5-6,   5.43 &  -0.0005  & 0.0059 \\
 6-7,   6.43 &   0.0058  & 0.0089 \\
 7-8,   7.44 &   0.0034  & 0.0132 \\
 8-10,  8.79 &   0.0077  & 0.0152 \\
10-12, 10.81 &  -0.0181  & 0.0282 \\
\end{tabular} \end{ruledtabular}
\end{table}

It has been confirmed with a ``bunch shuffling'' method
that the size of the statistical errors assigned is appropriate.
In this method,
the helicity of every beam bunch was newly assigned at random
and $\ALLcone$ was evaluated again.
Repeating this random assignment produced a large set of $\ALLcone$ values.
Its mean value should be of course zero and was confirmed in this exercise.
Its standard deviation indicates the size of the statistical fluctuation, 
and was consistent with the statistical errors assigned.
The point-to-point variance seems smaller than
the statistical errors of the data points,
but we could not find any unrecognized cause such as a statistical correlation.
We conclude that the small variance of the data points
happened statistically despite its small probability.

As a systematic error check,
the single spin asymmetry $A_L$ was measured.
It is defined as
\begin{equation}
 A_L \equiv 
    \frac{\sigma_+ - \sigma_-}{\sigma_+ + \sigma_-}
  = \frac{1}{P}\frac{N_+ - R\ N_-}{N_+ + R\ N_-}
 \ \ , \ \ \ 
 R \equiv \frac{L_+}{L_-},
\end{equation}
where
$N_{+}$ and $N_{-}$ are reconstructed-jet yields with one colliding proton beam having the positive and negative helicity, respectively;
$P$ is the beam polarization;
$R$ is the relative luminosity, i.e.~the ratio of the luminosity with 
the positive helicity ($L_+$) to that with the negative helicity ($L_-$).
As the jets are produced via the strong force, 
$A_L$ must be zero under the parity symmetry.
Thus any non-zero value indicates systematic errors.

\begin{figure}[bthp] 
\includegraphics[width=1.0\linewidth]{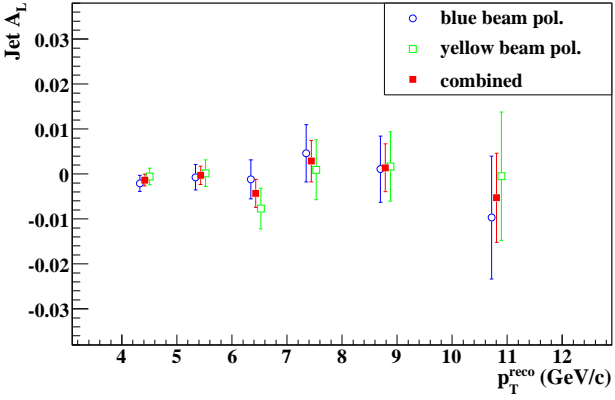}
\caption{ (color online) 
 Jet $A_L$ as a function of $\pTreco$.
 The blue and green points are the results using the polarizations
 of the blue beam and the yellow beam, respectively.
 The red points are the averages of the blue and green points.
}
\label{fig:A_L}
\end{figure}

Figure \ref{fig:A_L} shows measured $A_L$.
$A_L$ was measured for the polarization of one colliding beam while
the other beam was assumed to be unpolarized.
No significant asymmetry was observed.

\subsubsection{Constraint on $\Delta G$}

\begin{figure}[bthp] 
\includegraphics[width=1.0\linewidth]{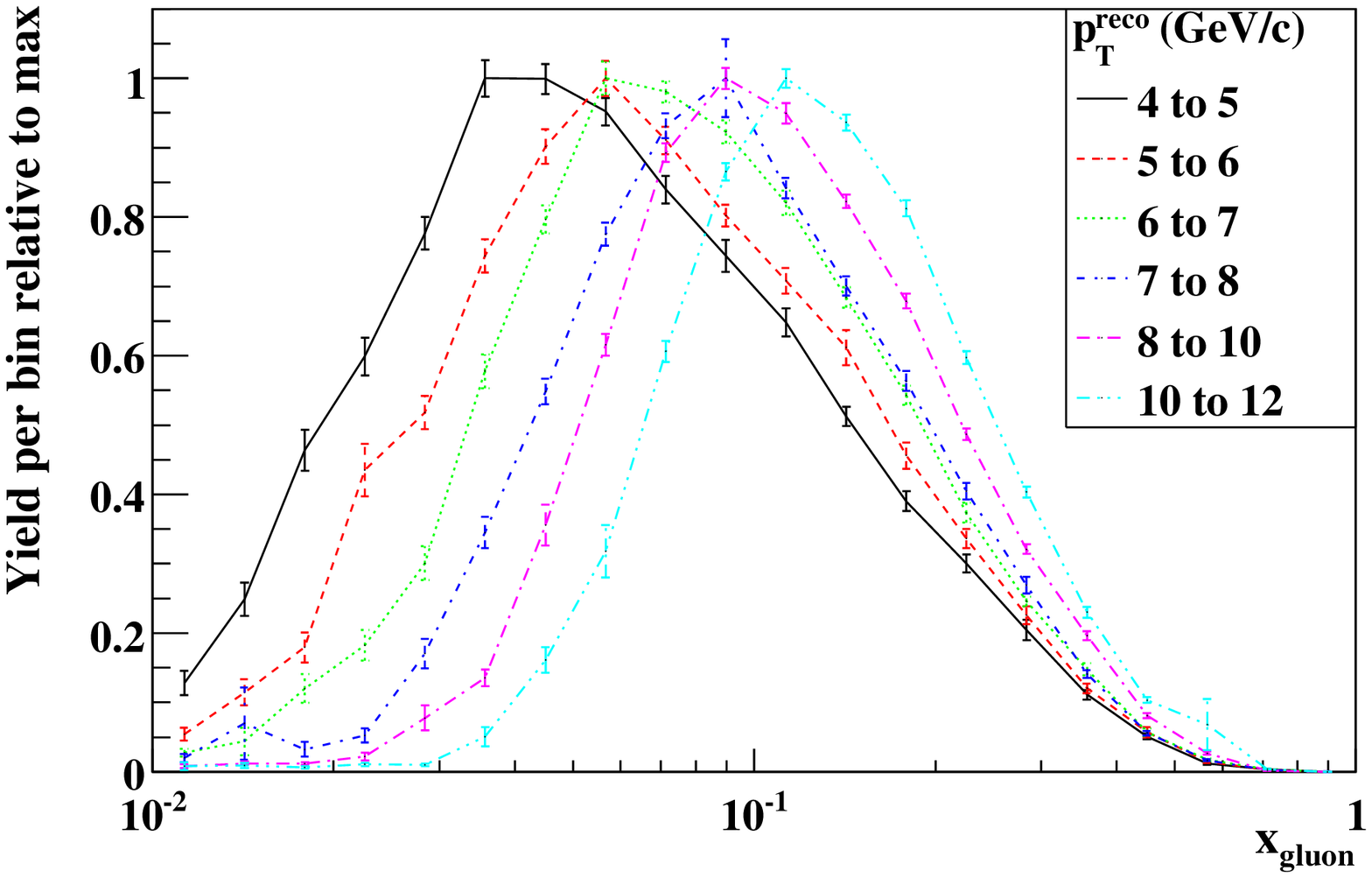}
\caption{ (color online) 
 Distributions of $x_{gluon}$ in events that include a reconstructed jet
 with $4 < \pTreco < 12$ GeV/$c$.
}
\label{fig:logxbj_dist}
\vspace{0.5cm}

\includegraphics[width=1.0\linewidth]{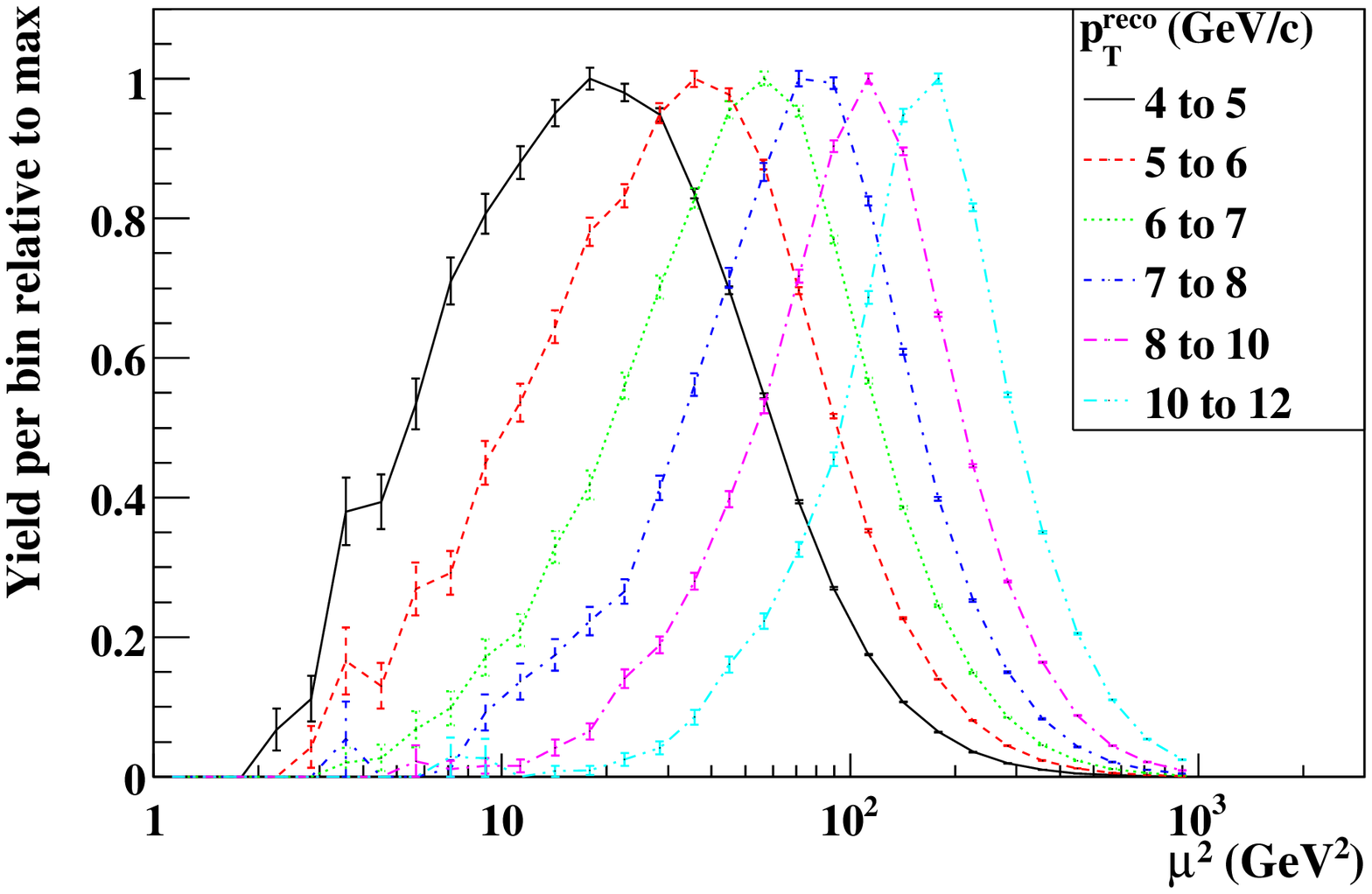}
\caption{ (color online) 
 Distributions of $\mu^2$ in events that include a reconstructed jet
 with $4 < \pTreco < 12$ GeV/$c$.
}
\label{fig:q2_dist}
\end{figure}

To determine the range of $x_{gluon}$ probed by this measurement,
the {\sc pythia} MPI simulation without {\sc geant} was used to
obtain event-by-event $x_{gluon}$ 
(one value per q-g scattering event, two values per g-g,
or none per q-q) and also $\mu^2$.
Figure \ref{fig:logxbj_dist} and \ref{fig:q2_dist}
show the distributions of $x_{gluon}$ and $\mu^2$, respectively.
The $x_{gluon}$ value where the yield is half maximum is
0.02 at the lower side of the ``$4 < \pTreco < 5$''
distribution and 0.3 at the upper side of the ``$10 < \pTreco < 12$''
distribution.
Therefore we adopt a range of $0.02 < x_{gluon} < 0.3$ as the range
probed by this measurement.
Table \ref{tab:delta_g_inte} shows the integral of $\Delta G(x)$
at the measured $x_{gluon}$ range, below the range and above the range.
The measured $x_{gluon}$ range includes $\sim$70\% of distributions
in all the four GRSV models shown.
With the same procedure, 
the $\mu^2$ range probed was estimated to be $5 < \mu^2 < 300$ GeV$^2$.

\begin{table}[bthp] 
\caption{ Partial integral of $\Delta G(x)$ at $\mu^2 = 1$ GeV$^2$. }
\label{tab:delta_g_inte}
\begin{ruledtabular} \begin{tabular}{lllll}
\multicolumn{1}{c}{Model} 
      & \multicolumn{4}{c}{$\int dx \Delta G(x)$ at each $x$ range} \\
      &  10$^{-4}$-0.02 & 0.02-0.3 & 0.3-1 & 10$^{-4}$-1 \\
\hline
$\Delta G=-G$ input & -0.406   & -1.09   & -0.208   &  -1.71   \\
                    & (24\%) & (64\%) & (12\%) &  \\
$\Delta G=0$ input  &  0.00808 &  0.0644 &  0.00869 &   0.0812 \\
                    & (10\%) & (79\%) & (11\%) &  \\
GRSV-std            &  0.0684  &  0.258  &  0.102   &   0.427  \\
                    & (16\%) & (60\%) & (24\%) &  \\
$\Delta G=G$ input  &  0.427   &  1.22   &  0.226   &   1.87   \\
                    & (23\%) & (65\%) & (12\%) &  \\
\end{tabular} \end{ruledtabular}
\end{table}

\begin{figure}[bthp] 
\includegraphics[width=1.0\linewidth]{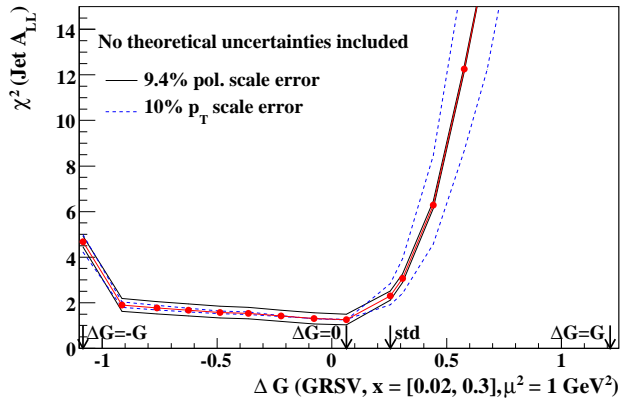}
\caption{ (color online) 
 $\chi^2$ between the measured $\ALL$ and the calculated $\ALL$
 as a function of the integrated value of $\Delta G(x)$.
}
\label{fig:chi2_vs_delta_g}
\end{figure}

Figure \ref{fig:chi2_vs_delta_g} 
shows the $\chi^2$ between the 6 data points
and the prediction curves as a function of the integral
$\int_{0.02}^{0.3} dx \Delta G(x, \mu^2=1)$ for each prediction curve.
The value of $\mu^2$ ($=1$ GeV$^2$) has been arbitrarily chosen in order
to show the value of the $\Delta G$ integral in horizontal axis.
Actual $\mu^2$ used in the $\ALL$ calculation varies depending on jet $\pT$.


The minimum of the $\chi^2$ is $\sim 1.5$ at $\Delta G = 0.07$,
namely the GRSV $\Delta G = 0$ input.  The 95\% and 99\% confidence
limits are where the $\chi^2$ increases from the minimum by 4 and 9,
respectively.  We obtained
\begin{equation}
-1.1 < \int_{0.02}^{0.3} \Delta G^{GRSV}(x, \mu^2=1) < 0.4 \\
\end{equation}
at 95\% confidence level and
\begin{equation}
\int_{0.02}^{0.3} \Delta G^{GRSV}(x, \mu^2=1) < 0.5 \\
\end{equation}
at 99\% confidence level.
In the assumptions of the present approach,
the error correlations between the normalization parameter and the
shape parameters in $\Delta G(x)$ are not included.
Also the fact that the shape of the polarized PDFs is 
parameterized into Eq.~\ref{eq:grsv_param} may cause 
additional uncertainty in $\Delta G(x)$.

\section{Conclusion}

We measured the event structure and the double helicity asymmetry ($\ALL$)
in jet production at midrapidity ($|\eta| < 0.35$) in longitudinally
polarized $p+p$ collisions at $\sqrt{s} = 200$ GeV were measured.
The main motivation is to use this complementary approach to inclusive 
measurements to better understand 
the contribution of the gluon spin ($\Delta G$) to the proton spin.
Because this measurement of $A_{LL}$ observes a larger fraction 
of the jet momentum, it reaches higher $p_T$ and thus higher gluon $x$.

The MPI-enhanced {\sc pythia} simulation agrees well with the real data 
in terms of the event structure: 
the multiplicity of photons and charged particles,
the $\pT$ density as a function of the azimuthal angle from
trigger photon, and the thrust in the PHENIX Central Arm.
A small difference in the intra-jet structure,
namely the fractions of photons and charged particles in jets,
was observed as shown in Fig.~\ref{fig:multiplicity}(c) to (f).
Nevertheless, the simulation well reproduces
the shape of jets and the underlying event at this collision energy.

In the measurement of jet $\ALL$,
measured particles were clustered by the seed-cone algorithm with a
cone radius $R=0.3$.
The relation between $\pTnlo$ and $\pTreco$ was evaluated with 
{\sc pythia} and {\sc geant}.
The jet production rate was measured and
satisfactorily reproduced by the calculation based on
the NLO pQCD jet production cross section and the simulation.
The jet $\ALL$ was measured at $4 < \pTreco < 12$ GeV/$c$. 
The main systematic errors are a $\pT$ scale uncertainty of 10\% and a
beam polarization uncertainty of 9.4\%.
The $x_{gluon}$ range probed by this jet measurement
with $4 < \pTreco < 12$ GeV/$c$ is mainly $0.02 < x < 0.3$ according to
the simulation.
The measured $\ALL$ was compared with the predicted values based on the
GRSV parameterization, and the comparison imposed the limit
$-1.1 < \int_{0.02}^{0.3}dx \Delta G^{GRSV}(x, \mu^2 = 1) < 0.4$
at 95\% confidence level or
$\int_{0.02}^{0.3}dx \Delta G^{GRSV}(x, \mu^2 = 1) < 0.5$
at 99\% confidence level.
The theoretical uncertainties such as 
the parameterization of the polarized PDFs
were not included in this evaluation.


\ 

\section*{ACKNOWLEDGMENTS}

We thank the staff of the Collider-Accelerator and Physics
Departments at Brookhaven National Laboratory and the staff of
the other PHENIX participating institutions for their vital
contributions.   
We also thank Werner Vogelsang for helpful 
discussions and calculations.  
We acknowledge support from the Office of Nuclear Physics in 
the Office of Science of the Department of Energy, the National 
Science Foundation, Abilene Christian University Research 
Council, Research Foundation of SUNY, and Dean of the College 
of Arts and Sciences, Vanderbilt University (U.S.A),
Ministry of Education, Culture, Sports, Science, and Technology
and the Japan Society for the Promotion of Science (Japan),
Conselho Nacional de Desenvolvimento Cient\'{\i}fico e
Tecnol{\'o}gico and Funda\c c{\~a}o de Amparo {\`a} Pesquisa do
Estado de S{\~a}o Paulo (Brazil),
Natural Science Foundation of China (People's Republic of China),
Ministry of Education, Youth and Sports (Czech Republic),
Centre National de la Recherche Scientifique, Commissariat
{\`a} l'{\'E}nergie Atomique, and Institut National de Physique
Nucl{\'e}aire et de Physique des Particules (France),
Ministry of Industry, Science and Tekhnologies,
Bundesministerium f\"ur Bildung und Forschung, Deutscher
Akademischer Austausch Dienst, and Alexander von Humboldt Stiftung (Germany),
Hungarian National Science Fund, OTKA (Hungary), 
Department of Atomic Energy (India), 
Israel Science Foundation (Israel), 
National Research Foundation and WCU program of the 
Ministry Education Science and Technology (Korea),
Ministry of Education and Science, Russia Academy of Sciences,
Federal Agency of Atomic Energy (Russia),
VR and the Wallenberg Foundation (Sweden), 
the U.S. Civilian Research and Development Foundation for the
Independent States of the Former Soviet Union, the US-Hungarian
NSF-OTKA-MTA, and the US-Israel Binational Science Foundation.



\end{document}